\documentclass[fleqn,usenatbib]{mnras}
\usepackage{newtxtext,newtxmath}
\usepackage[T1]{fontenc}

\DeclareRobustCommand{\VAN}[3]{#2}
\let\VANthebibliography\thebibliography
\def\thebibliography{\DeclareRobustCommand{\VAN}[3]{##3}\VANthebibliography}

\usepackage{anyfontsize}
\usepackage{graphicx}	% Including figure files
\usepackage{amsmath}	% Advanced maths commands
\usepackage{orcidlink}
\usepackage[flushleft]{threeparttable}
\usepackage{booktabs}
\usepackage{multirow}
\usepackage{array}
\usepackage[normalem]{ulem}
\usepackage{multicol}
\newcommand{\aref}[1]{\hyperref[#1]{Appendix~\ref*{#1}}}

\usepackage{xcolor}

\newcommand{\um}{{\mu\rm{m}}}

\newcommand{\km}{{\rm{km}}}
\newcommand{\au}{{\rm{au}}}
\newcommand{\gcc}{{{\rm g\ cm^{-3}}}}
\newcommand{\kms}{{{\rm km\ s^{-1}}}}
\newcommand{\Jkg}{{{\rm J\ kg^{-1}}}}
\newcommand{\Mearth}{{\rm M_{\earth}}}

\newcommand{\frho}{{f_{\rho}}}
\newcommand{\macc}{{m_{\rm{acc}}}}
\newcommand{\Mdisk}{{M_{\rm{d}}}}
\newcommand{\Dmin}{{D_{\rm{min}}}}
\newcommand{\Dmax}{{D_{\rm{max}}}}
\newcommand{\Dbl}{{D_{\rm{bl}}}}
\newcommand{\tcol}{{t_{\rm{c}}}}
\newcommand{\Qdstar}{{Q_{\rm{d}}^*}}

\newcommand{\vrel}{{v_{\rm{rel}}}}
\newcommand{\vim}{{v_{\rm{imp}}}}
\newcommand{\vesc}{{v_{\rm{esc}}}}

\newcommand{\vinf}{{v_{\infty}}}
\newcommand{\vsw}{{v_{\rm{SW}}}}

\newcommand{\tper}{{t_{\rm{per}}}}
\newcommand{\vk}{{v_{\rm{K}}}}
\newcommand{\taueff}{{\tau_{\rm{eff}}}}

\newcommand{\Dminlim}{{D_{\rm{KK}}}}

\newcommand{\Adisk}{{A_{\rm{disk}}}}
\newcommand{\Adust}{{A_{\rm{dust}}}}
\newcommand{\Adustgeom}{{A_{\rm{dust,geom}}}}
\newcommand{\Adustpr}{{A_{\rm{dust,PR}}}}

\newcommand{\fexppr}{{f_{\rm{exp,PR}}}}

\newcommand{\Ananodustgeom}{{A^{\prime}_{\rm{dust,geom}}}}
\newcommand{\delaprime}{{\Delta a^{\prime}}}
\newcommand{\Athindisk}{{A^{\prime}_{\rm{disk}}}}
\newcommand{\Adustprthin}{{A^{\prime}_{\rm{dust,PR}}}}
\newcommand{\Adustswthin}{{A^{\prime}_{\rm{dust,SW}}}}
\newcommand{\fexpprthin}{{f^{\prime}_{\rm{exp,PR}}}}
\newcommand{\fexpswthin}{{f^{\prime}_{\rm{exp,SW}}}}

\newcommand{\tpr}{{t_{\rm{PR}}}}
\newcommand{\Qpr}{{Q_{\rm{PR}}}}
\newcommand{\Qsw}{{Q_{\rm{SW}}}}
\newcommand{\tsw}{{t_{\rm{SW}}}}
\newcommand{\tem}{{t_{\rm{EM}}}}
\newcommand{\tsput}{{t_{\rm{sput}}}}
\newcommand{\tvisc}{{t_{\rm{visc}}}}
\newcommand{\tclr}{{t_{\rm{clear}}}}
\newcommand{\tclrpr}{{t_{\rm{clear,PR}}}}
\newcommand{\tclrsw}{{t_{\rm{clear,SW}}}}
\newcommand{\tclrem}{{t_{\rm{clear,EM}}}}
\newcommand{\tclrsput}{{t_{\rm{clear,sput}}}}
\newcommand{\Mdot}{{\dot{M}_{\star}}}
\newcommand{\Mdotsolar}{{\dot{\rm M}_{\sun}}}

\newcommand{\Mdiskdotpr}{{\dot{\Mdisk}_{\rm{,PR}}}}
\newcommand{\mdotsput}{{\dot{M}_{\rm{sput}}}}
\newcommand{\Ytot}{{Y_{\rm{tot}}}}

\newcommand{\nsw}{{n_{\rm{SW}}}}
\newcommand{\matom}{{m_{\rm{A}}}}

\newcommand{\mproton}{{m_{\rm{proton}}}}

\title[The fate of Mercury's ejected mantle]{The fate of Mercury's ejected mantle}

\author[T. Ghosh et al.]{
Tuhin Ghosh \orcidlink{0000-0002-3103-2000}$^{1}$\thanks{E-mail:\href{mailto:tg536@ast.cam.ac.uk}{tg536@ast.cam.ac.uk}},
Mark Wyatt \orcidlink{0000-0001-9064-5598}$^{1}$,
Alan P. Jackson \orcidlink{0000-0003-4393-9520}$^{2}$,
Oliver Shorttle \orcidlink{0000-0002-8713-1446}$^{1,3}$,
and Maximilian Sommer \orcidlink{0000-0003-4761-5785}$^{1}$
\\
$^{1}$Institute of Astronomy, University of Cambridge, Madingley Road, Cambridge CB3 0HA, UK\\
$^{2}$Department of Physics, Astronomy, \& Geosciences, Towson University, Towson, MD 21252, USA\\
$^{3}$Department of Earth Sciences, University of Cambridge, Downing Street, Cambridge CB2 3EQ, UK
}

\date{Accepted 2026 November 23. Received 2026 October 8; in original form 2026 June 24}

\pubyear{\the\year{}}

\begin{document}
\label{firstpage}
\pagerange{\pageref{firstpage}--\pageref{lastpage}}
\maketitle

%%%%%%%%%%%%%%%%%%%%%%%%%%%%%%%%%%%%%%%%%%%%%%%%%%
\begin{abstract}
Mercury’s unusually high core mass fraction, inferred from its high bulk density, remains difficult to explain. Mantle stripping by a giant impact provides a plausible origin, but without efficient removal, the ejected debris remains on Mercury-crossing orbits and later gets re-accreted. Here, we show that re-accretion of impact debris onto Mercury is inefficient, making a giant-impact origin plausible. We combine $N$-body simulations of giant-impact ejecta with a collisional evolution model and consideration of processes acting on small dust. We find that, when the ejecta are dominated by large, several hundred km-sized non-vaporized fragments, the debris disk remains optically thin and the collisional cascade is truncated near the radiation-pressure blowout limit, where micron-sized grains are removed. In contrast, if vaporized, the ejecta recondense into mm to cm-sized droplets and form an optically thick disk, in which radiation pressure cannot efficiently truncate the cascade. The debris can then be ground down to nanometre-sized grains, where collisions become predominantly non-fragmenting and dissipative rather than destructive. Subsequent inelastic collisions damp the eccentricities and inclinations of the nanodust, producing a geometrically thin disk whose lifetime is governed by removal through exposed surface layers, where grains acquire large charge-to-mass ratios through solar UV radiation and are rapidly removed by electromagnetic forces. In either scenario, collisional evolution strongly reduces the re-accretion of debris and alleviates a key difficulty for the giant-impact origin of Mercury’s large core. More broadly, highly energetic giant impacts in the inner regions of planetary systems may lead to short-lived populations of nanometre-sized dust.
\end{abstract}


%%%%%%%%%%%%%%%%%%%%%%%%%%%%%%%%%%%%%%%%%%%%%%%%%%

%%%%%%%%%%%%%%%%% BODY OF PAPER %%%%%%%%%%%%%%%%%%
\section{Introduction}
%%%%%%%%%%%%%%%%%%%%%%%%%%%%%%%%%%%%%%%%%%%%%%%%%%

Mercury’s bulk density, $5.43\, \gcc$, is unusually high for its mass, being comparable to Earth’s despite the planet being $\sim20$ times less massive and therefore subject to much weaker gravitational self-compression \citep[e.g.,][]{Ash_1971, Anderson_1987}. This high density implies a disproportionately large metallic core comprising $\sim70\%$ of the planet’s mass -- roughly twice the core mass fraction (CMF) of Earth and other terrestrial planets \citep[e.g.,][]{Harder_Schubert_2001,Solomon_2003, Hauck_2013}. The origin of this extreme metal enrichment is not yet fully understood.

Several hypotheses have been proposed to explain Mercury’s anomalously large metal core, which fall into two broad classes. In one class of models, Mercury’s high CMF is primordial, with the planet forming from intrinsically iron-rich material produced by chemical, magnetic or gas-drag related fractionation in the inner protoplanetary disk \citep[e.g.,][]{Lewis_1972, Weidenschilling_1978, Hubbard_2014, Kruss_Wurm_2018, Kruss_Wurm_2020, Johansen_Dorn_2022}. In the other, Mercury forms with a more Earth-like composition, but after differentiation subsequently loses a large fraction of its silicate mantle through an energetic giant impact \citep[e.g.,][]{Benz_1988, Benz_2007, Cameron_1988, Chau_2018}. Giant impacts are a natural outcome of late-stage terrestrial planet formation, and are thought to have shaped several features of the Solar System, including the formation of the Moon \citep[e.g.,][]{Hartmann_1975, Cameron_Ward_1976, 1986_Benz, 2001_Canup, Cuk_Stewart_2012, Lock_2018, Kegerreis_2022}, Martian moons \citep[e.g.,][]{Citron_2015, Canup_Salmon_2018} and the large obliquity of Uranus \citep[e.g.,][]{Safronov_1966, Lissauer_1991, Morbidelli_2012,Kegerreis_2018}, making the giant impact hypothesis for the origin of Mercury's high density particularly appealing. However, a key difficulty in this scenario is the fate of the ejected material. A substantial fraction of the stripped mantle is expected to remain on Mercury-crossing orbits and be re-accreted on Myr timescales, thereby reducing the planet’s core mass fraction and partially undoing the mantle-stripping effect of the giant impact \citep{Benz_2007,Gladman_Coffey_2009}. More recently, \citet{Scora_2024} found that high-CMF Mercury analogues can form during terrestrial planet formation, but subsequent accretion of impact debris reduces their final CMF, further highlighting the importance of determining whether the stripped material is permanently removed.

Various mechanisms have been invoked that act on the ejected material and limit re-accretion, including Poynting–Robertson (PR) drag \citep[e.g.,][]{Benz_1988, Benz_2007} and enhanced solar wind (SW) drag in the early Solar System \citep{Spalding_Adams_2020}. These processes can transport small particles inward or expel them from the system, but their efficiency is highly sensitive to particle size and the optical depth of the debris cloud, which may be large for a massive debris disk in the inner Solar System, and their ability to deplete the bulk of the ejecta remains uncertain.

However, previous studies have largely focused on the dynamical evolution of ejected debris, without accounting for the collisional evolution of the debris population or the effect of optical depth on dust removal. A massive debris cloud at sub-au distances is expected to collide rapidly, producing progressively smaller dust particles that are subject to different, and sometimes much faster, removal processes \citep[e.g.,][]{Wyatt_2008}. Collisional evolution can thereby alter both removal and re-accretion rates \citep{Ghosh_2025}. Yet the role of collisional evolution, together with the optical-depth constraints on small-grain removal, has not been systematically explored for Mercury's ejecta.

In this study, we revisit the fate of Mercury's ejected mantle using $N$-body dynamical models coupled with collisional evolution of the debris. Specifically, we assess whether collisional grinding can efficiently deplete the debris reservoir and thereby suppress re-accretion. The numerical setup and assumptions are described in \autoref{sec:setup}. Our key results are presented in \autoref{sec:res}, followed by a discussion of their implications and limitations in \autoref{sec:discussion}. We summarize our conclusions in \autoref{sec:conclusions}.

%%%%%%%%%%%%%%%%%%%%%%%%%%%%%%%%%%%%%%%%%%%%%%%%%%
\section{Numerical Setup}\label{sec:setup}
%%%%%%%%%%%%%%%%%%%%%%%%%%%%%%%%%%%%%%%%%%%%%%%%%%

\begin{figure*}
\includegraphics[width=\textwidth]{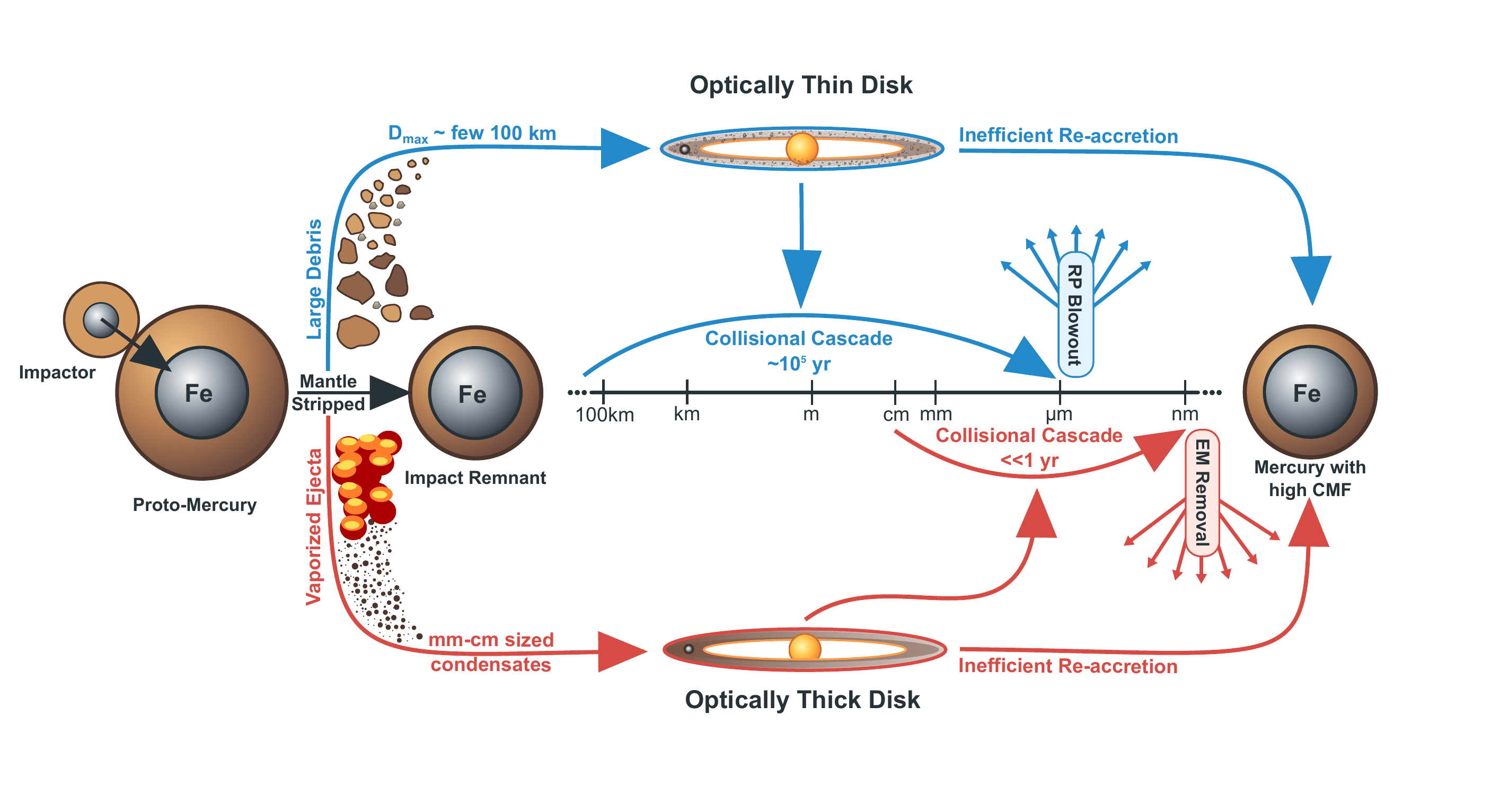}
\vspace{-25pt}
\caption{A schematic diagram showing the two distinct evolutionary pathways of Mercury's giant-impact ejecta, depending on the physical state of the ejecta after impact. Blue arrows show the collisional evolution of large, non-vaporized debris fragments, discussed in \autoref{subsec:large_debris}, while red arrows show that of vaporized ejecta, discussed in \autoref{subsec:vaporized_debris}. RP denotes radiation pressure, EM electromagnetic removal, and CMF core mass fraction.}
\label{fig:summary}
\end{figure*}

In this study, we focus on the evolution of debris ejected during a giant impact event responsible for stripping Mercury’s mantle. In the canonical scenario, a proto-Mercury with $\sim 2.25$ times its present mass is impacted by a body with mass $\sim 1/6$ that of the target, ejecting $\sim 0.1\,\Mearth$ of debris \citep[e.g.,][]{Benz_1988,Benz_2007}. Our aim is not to reproduce the detailed ejecta distribution of a specific impact event, but to test whether re-accretion remains significant once collisional evolution is included.

Key to the approach we adopt, therefore, is to simplify the impact event itself in order to follow the collisional and re-accretion processes, and in making simplifications to make choices that maximise re-accretion onto the planet: in this way, our simulations are conservative about the efficacy of giant impacts for stripping Mercury's mantle. The simplified framework adopted here is designed to follow two distinct physical outcomes for the ejecta, depending on whether the stripped mantle remains in large fragments or is vaporized and recondenses into small droplets. These two pathways are summarized schematically in \autoref{fig:summary}. The total mass of the ejecta is treated as a free parameter and varied over $0.02$--$0.5\,\Mearth$, encompassing the canonical estimate and accounting for uncertainties in impact geometries, progenitor mass, and outcomes. The ejecta are represented by $40{,}000$ non-interacting test particles, initially distributed isotropically at a distance of $50$ planetary radii from Mercury. Each particle is assigned an ejection velocity corresponding to $\vinf = 4\,\kms$.

High-resolution hydrodynamic impact simulations of Mercury's mantle-stripping event will produce ejecta distributions that differ from the idealized initial conditions adopted here. In particular, impact ejecta are not expected to be launched isotropically or with a single velocity, but instead with a geometry and velocity distribution that depend sensitively on the specific impact conditions.
For the subsequent dynamical evolution and re-accretion considered here, the sensitivity to these initial conditions primarily arises from the direction of ejection relative to Mercury's orbital motion and the ejecta velocity. We test the sensitivity to the former in Section 3, while the dependence on ejection velocity has previously been explored by \citet{Gladman_Coffey_2009}.
For Mercury, the collision must be sufficiently energetic to remove a large fraction of the mantle, implying an impact velocity $\vim \gtrsim 2\vesc$, where $\vesc$ is the escape velocity from the planet’s surface ($\vesc = 4.25\ \kms$ for present-day Mercury). Given that $\vinf \sim \vim/2$ \citep{Emsenhuber2024}, we expect $\vinf \gtrsim 4\ \kms$. Moreover, \citet{Gladman_Coffey_2009} found that re-accretion onto Mercury is significantly higher at $\vinf = 4\ \kms$ than at larger ejection velocities, as higher $\vinf$ disperses the debris over a wider annulus, thereby reducing re-accretion efficiency. This motivates our adoption of $\vinf = 4\ \kms$ to provide a conservative assessment of re-accretion efficiency. Larger $\vinf$ values, as may arise in detailed giant-impact simulations, are therefore expected to reduce, rather than enhance, re-accretion: if re-accretion remains inefficient in the simplified setup adopted here, the detailed ejecta properties of a specific impact simulation are unlikely to alter our qualitative conclusions. The assumed isotropic initialization of debris ejection directions provides an impact-geometry-independent starting condition for our dynamical simulations. We revisit this assumption and test the sensitivity of our results to the ejection direction in \autoref{sec:res}.

We evolve these particles for $10\,\mathrm{Myr}$ under the gravitational influence of the Solar System planets from Mercury through Jupiter, assumed to occupy their present-day orbital configurations. Planets beyond Jupiter are excluded, as they are not expected to significantly perturb the debris, and no test particles in our simulations went beyond Jupiter’s orbit. We use the \texttt{MERCURIUS} hybrid integration scheme \citep{2019REBMERCURIUS}, implemented in the \texttt{REBOUND} simulation package \citep{ReinREB2012}, with an initial timestep of $2$ days. Particles are removed upon collision with a planet or the Sun, or if they reach a heliocentric distance beyond $50\,\au$, at which point they are considered ejected.

In addition to the dynamical evolution under the gravitational influence of the planets, the debris population also evolves through mutual collisions. Frequent collisions among debris particles lead to a collisional cascade, progressively fragmenting material to smaller sizes and ultimately producing sub-micron sized and smaller dust that can be efficiently removed by various processes \citep[e.g.,][]{Wyatt_2008}.
We model the collisional evolution with a statistical approach using the outputs of our dynamical integrations. We treat the debris particles in the $N$-body simulations as tracer particles that sample the dynamically evolving spatial distribution and relative velocity field of the debris under the gravitational influence of the planets. We do not explicitly track collisional fragmentation in the $N$-body simulations. Instead, the collisional cascade is treated statistically.

We assume that the collisional fragments inherit similar orbital properties and evolve gravitationally following their parent population traced by the $N$-body particles. Each $N$-body particle is thus representative of both the debris created in the original giant impact, as well as the fragments created in subsequent collisions of that debris. As a result, our model does not capture collisional damping of eccentricities and inclinations. However, in the regime where collisions are predominantly fragmenting and occur at high relative velocities, collisional damping is expected to be minimal \citep{Jankovic_2024}.

We assume that the mass represented by each tracer particle follows a size distribution $n(D) \propto D^{-3.5}$, corresponding to a steady state collisional cascade \citep[e.g.,][]{Dohnanyi_1969,Wyatt_2008}. We estimate the local collision rates at the location of each tracer particle using the spatial distribution of the debris obtained from the $N$-body simulations. For each particle, we identify its $10$ nearest neighbours and use the volume enclosing these particles to estimate the local number density ($n$), normalized to the total debris mass, as well as the nearest neighbour relative velocity, $\vrel$. The collision rate for each particle is then computed using a particle-in-a-box approximation as $R = n\sigma_{\mathrm{cc}}\vrel$, where $\sigma_{\mathrm{cc}}$ is the collisional cross-section integrated over the adopted size distribution following \citet{Jackson_2014}. The debris mass is initially distributed equally among the tracer particles, and their masses are subsequently updated over time according to their local collision rates. This framework thus couples the dynamical and collisional evolution of the debris in a self-consistent manner, enabling us to quantify the mass re-accreted onto Mercury while accounting for collisional depletion.

%%%%%%%%%%%%%%%%%%%%%%%%%%%%%%%%%%%%%%%%%%%%%%%%%%
\section{Results}\label{sec:res}
%%%%%%%%%%%%%%%%%%%%%%%%%%%%%%%%%%%%%%%%%%%%%%%%%%

\begin{figure*}
    \centering
    % First Row
    \includegraphics[width=0.325\textwidth]{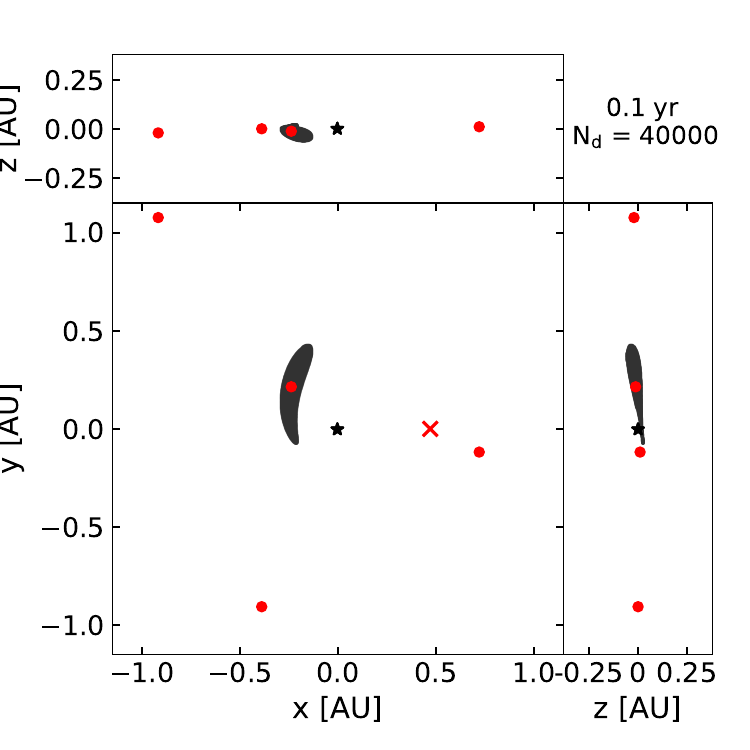}\hspace{0.1em}
    \includegraphics[width=0.325\textwidth]{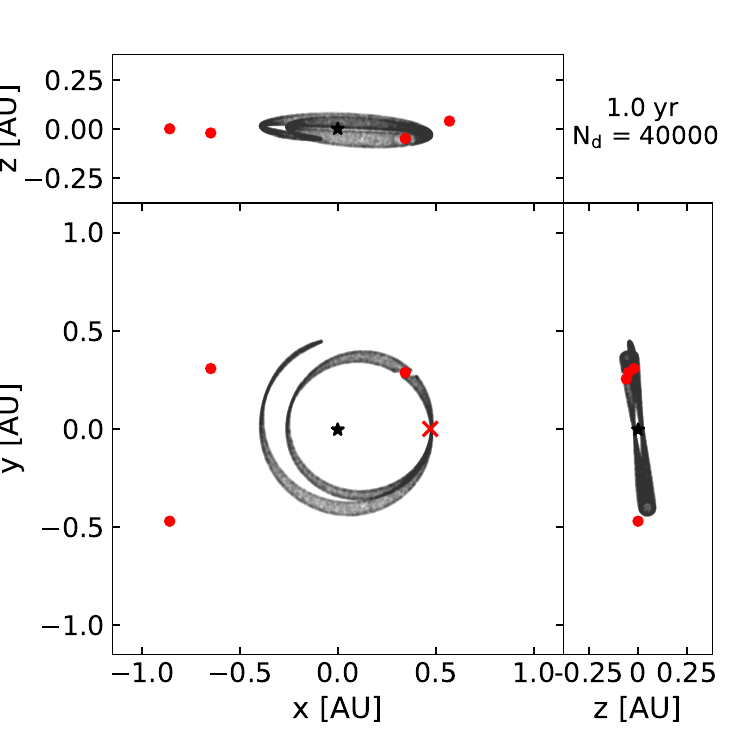}\hspace{0.1em}
    \includegraphics[width=0.325\textwidth]{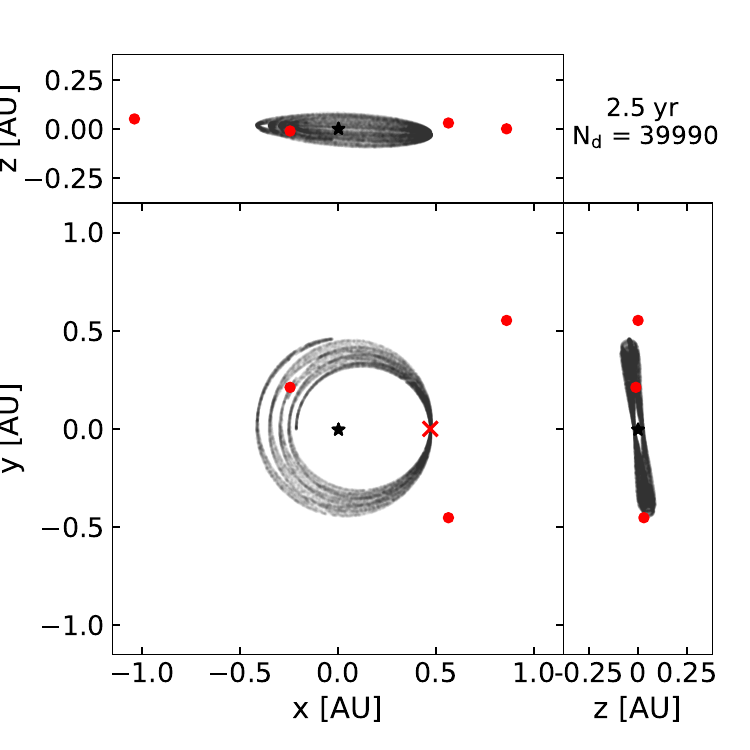}
    \vspace{1ex}
    % Second Row
    \includegraphics[width=0.325\textwidth]{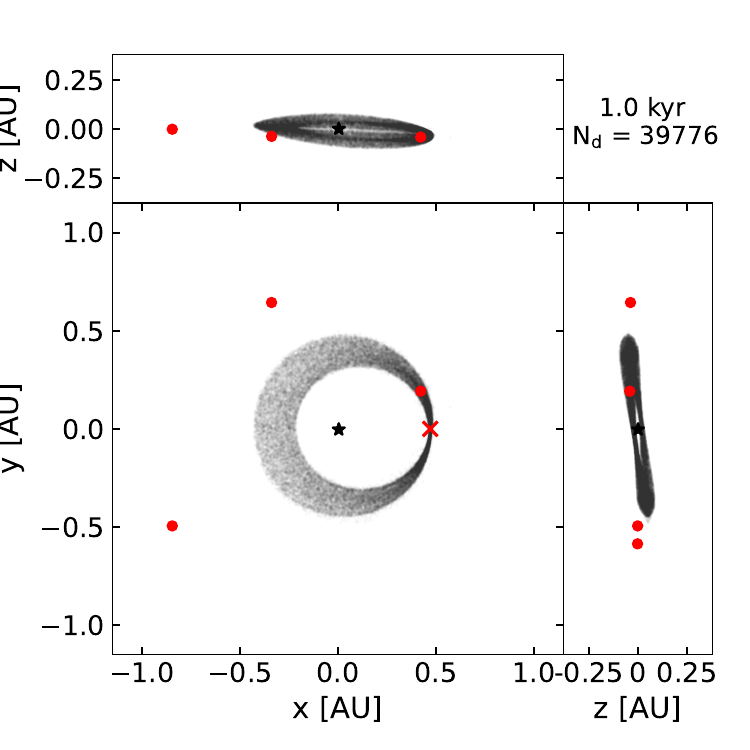}\hspace{0.1em}
    \includegraphics[width=0.325\textwidth]{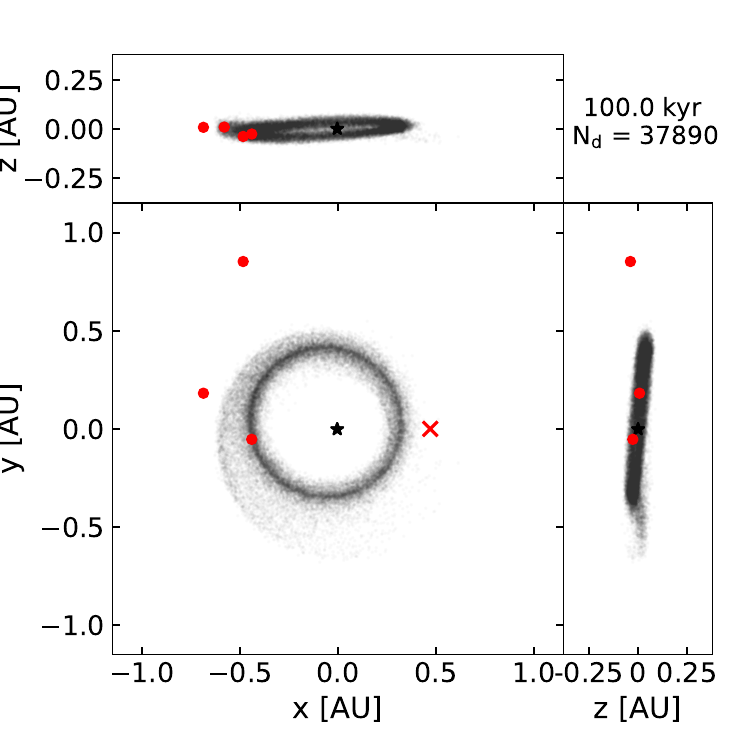}\hspace{0.1em}
    \includegraphics[width=0.325\textwidth]{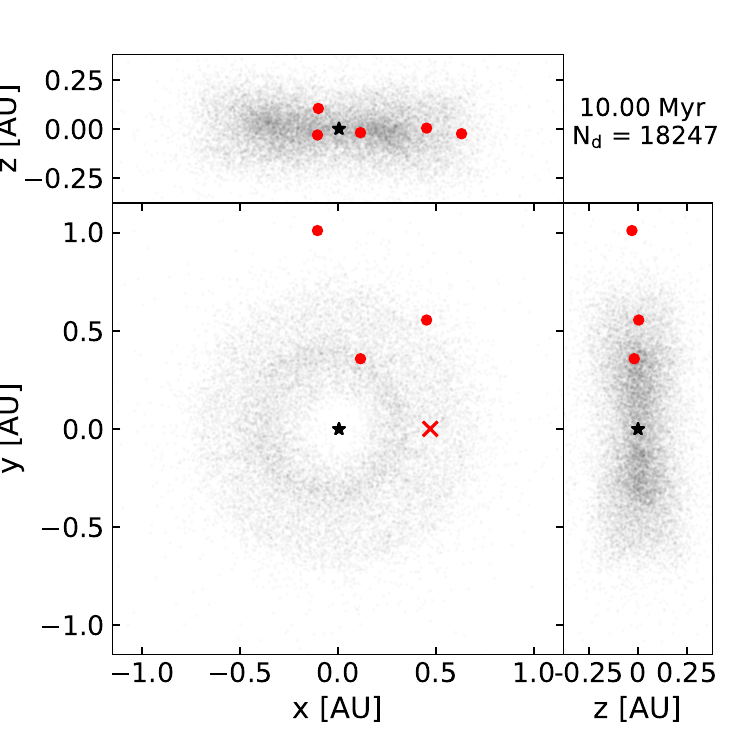}
    % Caption and label
    \caption{Snapshots of the evolution of debris ejected isotropically from Mercury with $\vinf = 4\ \kms$. Debris particles (black dots) and the innermost five planets (red circles) are shown in three orthogonal projections: the orbital plane ($x$–$y$, main panel) and the $x$–$z$ and $z$–$y$ planes (top and right panels, respectively). The red cross marks the location of the collision point. The simulation time and debris particle count $N_{\rm d}$, are indicated at the upper right corner of each panel. The $x$--$y$ plane is chosen to be the Solar System's invariable plane. The debris particles shown here are massless test particles and do not have physical sizes.}
    \label{fig:snapshots}
\end{figure*}

We first describe the purely dynamical evolution of Mercury's giant-impact ejecta, which sets the spatial distribution and relative velocity field that control the collisional evolution, as well as the re-accretion history. The snapshots of our $N$-body simulations at different stages are shown in \autoref{fig:snapshots}. The overall evolution closely follows that found in studies of impact-generated debris \citep{Jackson_Wyatt_2012, Jackson_2014}, albeit on timescales appropriate for the orbital properties in the vicinity of Mercury. In our simulations, Mercury has its present-day eccentricity of $e\simeq0.21$, and the impact occurs near the apocentre, where the debris is released. Immediately after release, the debris forms a tightly wound spiral structure as particles with orbital periods shorter than Mercury move ahead of the planet, while those on longer-period orbits trail behind. A key feature of this early evolution is that all debris orbits initially pass through a single point in space, referred to as the collision point, corresponding to the location of the impact. As a result, the debris distribution develops a strong asymmetry, with a pronounced density enhancement at the collision point and a corresponding feature along the anti-collision line on the opposite side of the orbit. Since all particles repeatedly pass through the collision point, this region is expected to experience an enhanced rate of mutual collisions compared to the rest of the disk. As the system evolves, the spiral structure becomes increasingly tight, and by $\sim 1\,\mathrm{kyr}$ the spiral structure weakens, transforming the debris disk into a continuous, but highly asymmetric, ring (bottom left panel of \autoref{fig:snapshots}). The density enhancement at the collision point persists for several kyr, but at later times ($\sim 10^5\,\mathrm{yr}$, bottom center panel of \autoref{fig:snapshots}), differential precession of the debris orbits, driven by continued perturbations by Mercury and other planets, gradually erases the associated asymmetry. The debris orbits inherit a finite eccentricity from Mercury's orbit, with an additional spread introduced by the ejecta velocity distribution. Over time, apsidal precession can shift their orbital orientations so that the bulk of the material no longer passes near the original collision point, as seen in the $10^5\,\mathrm{yr}$ snapshot. By this stage, a low-density, extended component is also evident. Unlike the dense inner component, which remains largely aligned with Mercury’s orbit, this extended population is more closely aligned with the invariable plane of the planetary system, suggesting the influence of secular perturbations from the other planets.

\begin{figure}
\includegraphics[width=\columnwidth]{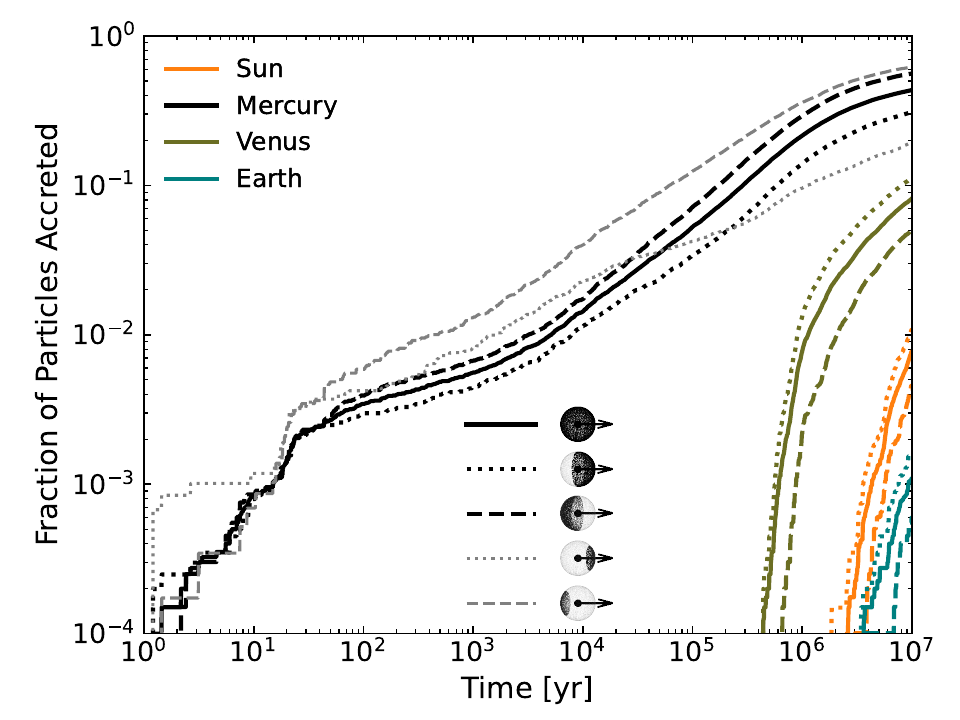}
\caption{Cumulative fraction of test particles ejected from Mercury that are later accreted by the Sun (orange), Mercury (black), Venus (dark olive) and Earth (teal) as a function of time. Solid lines correspond to the full isotropic ejecta distribution, while dotted and dashed lines correspond to the two launch hemispheres oriented along and opposite to Mercury’s orbital motion, respectively. The grey dotted and dashed lines show the particles re-accreted by Mercury from narrower $45^\circ$ launch cones about the same two directions.
}
\label{fig:debris_accretion}
\end{figure}
During the integration, we track the number of debris particles accreted by each planet, as shown in \autoref{fig:debris_accretion}. We find that about $43.5\%$ of the debris particles ($17{,}387$ out of $40{,}000$) are re-accreted onto Mercury within $10\,\mathrm{Myr}$. This is consistent with the results of \citet{Gladman_Coffey_2009}, whose simulations found that the cumulative re-accreted fraction reaches a comparable value by $10\,\mathrm{Myr}$ and continues to increase to about $50\%$ over $30\,\mathrm{Myr}$. Mercury’s nearest neighbour, Venus, accretes $\sim 8.2\%$ of the debris ($3279$ out of $40{,}000$), predominantly at later times ($\gtrsim 0.4\,\mathrm{Myr}$), once a low-density extended component of the disk develops and begins to populate Venus-crossing orbits (\autoref{fig:snapshots}). A small fraction of the particles, $310$ out of $40{,}000$ ($\sim0.8\%$), are lost to the Sun. Accretion onto the other planets is minimal, with Earth accreting only $44$ particles, and Mars only $1$ out of $40{,}000$ over the same interval. The early evolution is instead dominated by interactions with Mercury, since the bulk of the debris has orbits that pass through the collision point along with Mercury. Even at later times, Mercury remains embedded in the disk's dense component and thus continues to accrete material.

As a check on the sensitivity of our results to the idealized launch geometry, we split the initially isotropic ejecta into two hemispheres oriented along and opposite to Mercury’s orbital motion. Particles launched opposite to Mercury’s orbital velocity are re-accreted by Mercury more efficiently than those launched along the direction of orbital motion (\autoref{fig:debris_accretion}). Ejecta launched opposite to the orbital motion have lower heliocentric orbital energy and angular momentum, keeping them on more compact Mercury-crossing orbits. In contrast, ejecta launched along the orbital motion have higher energy and spread outward, where they can be scattered or accreted by Venus. Venus consequently accretes more efficiently from this population at late times. The isotropic case lies between these two limiting populations. An analogous split into particles launched toward and away from the Sun produces no significant difference in the accretion histories, indicating that the tangential component of the launch velocity is the dominant directional effect. For the orbital-direction split, the final re-accreted fraction onto Mercury varies between $31.0\%$ and $56.1\%$, compared to $43.5\%$ for the isotropic case. Restricting the same comparison to narrower $45^\circ$ launch cones gives a wider range, from $19.6\%$ to $62.0\%$, as expected for more strongly directional subsets. Even for this more restrictive split, the corresponding variation in the re-accreted debris fraction remains small compared to the factor of 25 range in initial disk mass ($0.02<\Mdisk/\Mearth<0.5$) explored in this study. In the remainder of this paper, we adopt the full isotropic ejecta distribution as the fiducial initial condition, since it preserves the angle-averaged behaviour of the idealized launch model and does not alter the key conclusion that a substantial fraction of the debris remains susceptible to re-accretion unless additional removal processes operate.

\begin{figure}
\includegraphics[width=\columnwidth]{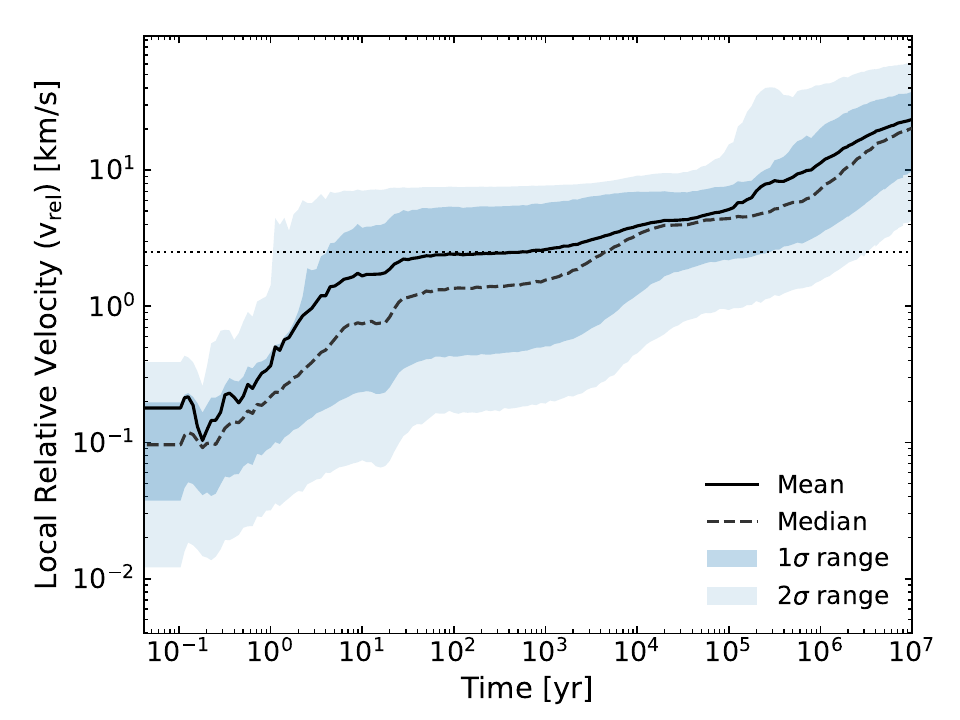}
\caption{Distribution of local relative velocities ($\vrel$) over time. The solid and dashed curves show the mean and median values, respectively, while the shaded regions indicate the $1\sigma$ (darker) and $2\sigma$ (lighter) ranges of the $\vrel$-distribution. The horizontal dotted line marks a reference velocity of $2.5\,\kms$.}
\label{fig:local_vrel}
\end{figure}
\autoref{fig:local_vrel} shows the evolution of the local relative velocity distribution. Immediately after ejection, particles have low relative velocities of $\sim 0.1\,\kms$, and the velocity distribution remains concentrated at low values for approximately one orbital period of Mercury as the isotropically ejected particles spread out. During this initial expansion, mutual collisions should be rare, as the particles are still moving away from the ejection point and have not yet re-encountered one another along their orbits. Gradually, particles with different semi-major axes move out of phase and randomize their mean anomalies. This increases $\vrel$ and allows particles to begin colliding with one another on crossing orbits. The distribution exhibits a large spread at all times, with velocities spanning over an order of magnitude, while both the mean, median, and the high-velocity tail shift to higher values as the system evolves. The mean relative velocity reaches $\sim 2.5\,\kms$ by $\sim 10$--$50\,\mathrm{yr}$ and remains at this level for $\sim 50$--$10^3\,\mathrm{yr}$ before increasing further up to $\sim \mathrm{few}\times 10\,\kms$ at $10\,\mathrm{Myr}$ as the disk continues to spread and is perturbed by the planets. These high relative velocities imply that debris collisions are highly energetic throughout the evolution and are expected to be predominantly fragmenting.

The dynamical evolution described above implies that debris particles experience high relative velocities, leading to frequent and energetic collisions. The outcome of this collisional evolution depends sensitively on the physical state of the ejecta, because this determines the initial size distribution and the optical depth of the debris disk. The physical state of the ejecta is set by the impact conditions. In relatively lower-energy impacts, a substantial fraction of the ejecta may remain in large fragments \citep[e.g.,][]{Genda_2015, Watt_2024}. In contrast, highly energetic impacts can vaporize much of the silicate mantle, with the vaporized ejecta subsequently cooling and condensing into much smaller droplets \citep[e.g.,][]{Benz_2007,Lock_2018}. We therefore consider two limiting scenarios: one in which Mercury's mantle is ejected as large debris (\autoref{subsec:large_debris}), and another in which it is largely vaporized and subsequently condenses into small particles (\autoref{subsec:vaporized_debris}).
These two distinct collisional-evolution pathways, summarized schematically in \autoref{fig:summary}, are explored in detail in the following sections.

% ------------------------------------------------------%
\subsection{Fate of large debris fragments}\label{subsec:large_debris}
% ------------------------------------------------------%

\begin{figure}
\includegraphics[width=\columnwidth]{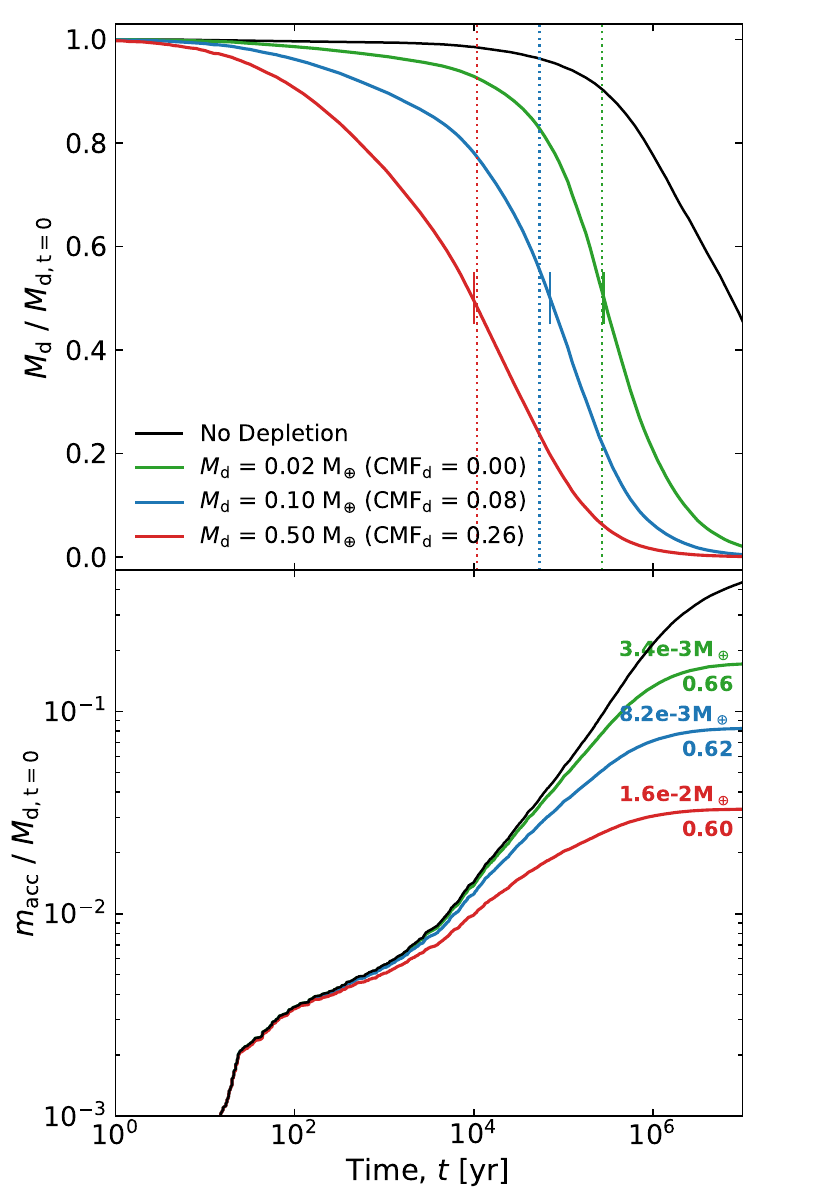}
\caption{Debris mass evolution (top) and fraction of the initial disk mass re-accreted by Mercury (bottom). Solid colored curves include collisional depletion for different debris masses, assuming a maximum debris fragment size of $\Dmax=600\,\km$, while the black curves show the purely dynamical case without collisional depletion. The final accreted masses and corresponding final CMFs of Mercury, after accounting for collisional depletion, are indicated in the bottom panel. Solid vertical bands in the top panel mark the numerically estimated collisional timescales, while dotted vertical lines show the corresponding analytic estimates from \autoref{eq:tcol}.}
\label{fig:coll_depletion}
\end{figure}
In this section, we consider the limiting case where material is ejected as large, non-vaporized bodies. The size distribution of debris produced in giant impacts is highly uncertain, but high-resolution Smoothed Particle Hydrodynamics (SPH) simulations suggest that the largest fragments can reach sizes of several hundred kilometres \citep[e.g.,][]{Genda_2015, Watt_2024}. We therefore adopt a maximum fragment size of $\Dmax = 600\,\km$\footnote{The non-vaporized ejecta are expected to span a broad range of sizes. At the upper end of the size distribution are large gravitationally bound clumps found in SPH simulations, which motivates our choice of $\Dmax$. The size distribution also extends to smaller sizes, including metre-scale or even smaller \citep[e.g.,][]{Hyodo_2017}. The whole distribution evolves through a collisional cascade, ultimately producing small dust grains that are subject to non-gravitational forces. These non-vaporized ejecta are distinct from the vaporized ejecta considered in \autoref{subsec:vaporized_debris}, which recondense into mm--cm-sized particles \citep[e.g.,][]{Melosh_Vickery_1991, Johnson_Melosh_2014}.}, consistent with these SPH results, and assign a catastrophic disruption threshold $\Qdstar \sim 1.2\times10^{6}\,\mathrm{J\,kg^{-1}}$ appropriate for bodies of this size \citep{Benz_Asphaug_1999}. Since most of the mass in a steady-state cascade with $n(D)\propto D^{-3.5}$ resides near $\Dmax$, the global depletion rate is mainly determined by the disruption of the largest bodies. For a fixed total debris mass, a larger $\Dmax$ implies lower collision rates \citep[see Eq.~A15 of][]{Ghosh_2025}. If the maximum fragment size were smaller, the debris would be depleted at a faster rate, leading to even lower re-accretion efficiencies than in the scenario considered here. Thus, the choice of a large $\Dmax$ favors re-accretion and provides a conservative estimate of the collisional depletion.

In a steady-state collisional cascade with the size distribution adopted here ($\propto D^{-3.5}$), the total geometric cross-section of the debris is dominated by the smallest grains and can be estimated as
\begin{equation}\label{eq:adust}
    \Adustgeom \approx \frac{3\Mdisk}{2\rho\sqrt{\Dmax \Dmin}},
\end{equation}
where $\rho$ is the debris density and $\Dmin$ is the minimum grain size in the cascade. In optically thin disks, $\Dmin$ is set by the radiation pressure blowout limit $\Dbl$, below which radiation pressure makes particles unbound, and they are blown out from the system on orbital timescales. For Solar-type stars, $\Dbl \sim 0.8\,\um$ \citep[e.g.,][]{Wyatt_2008}. In the large debris limit ($\Dmax \gtrsim 200\;\km$) Mercury's impact debris disk remains optically thin with $\Dmin = \Dbl$ even in the extreme case of $\Mdisk = 0.5\,\Mearth$. In this regime, $\Adustgeom < \Adisk \sim 2\pi a\,\Delta h$, where $a \sim 0.4\,\au$ is the orbital radius and $\Delta h \sim 0.074\,\au$ is the vertical extent of the disk inferred from the $99^\mathrm{th}$ percentile spread of particles relative to Mercury's orbital plane (which closely traces the debris midplane, see \autoref{fig:snapshots}) in our $N$-body simulations with $\vinf=4\,\kms$. Thus, in the optically thin large-fragment regime, the collisional cascade continuously supplies grains down to micrometre sizes, where they are directly exposed to solar radiation and rapidly blown out of the system by radiation pressure.

The collisional mass depletion of this optically thin debris disk and the corresponding mass re-accretion onto Mercury are shown in \autoref{fig:coll_depletion} and summarized in \autoref{tab:cmf_estimates}. The depletion timescale is sensitive to both the maximum fragment size and the initial disk mass. A larger $\Dmax$ increases the depletion timescale by placing more mass in large fragments, while a larger $\Mdisk$ increases the collision rate and leads to faster depletion. We adopt $\Mdisk=0.1\,\Mearth$ as our fiducial case, motivated by the canonical Mercury-stripping impact, and explore masses of $0.02$ and $0.5\,\Mearth$ to test the sensitivity to the total ejecta mass. In the absence of collisional depletion, the disk retains $\sim 45.6\%$ of its initial mass after $10\,\mathrm{Myr}$ (top panel), with most of the loss due to re-accretion onto Mercury (bottom panel, also see \autoref{fig:debris_accretion}). However, when collisional evolution is accounted for, the disk is efficiently depleted on sub-Myr timescales. For the canonical impact disk with $\Mdisk = 0.1\,\Mearth$, the disk is almost entirely depleted within $10\,\mathrm{Myr}$, resulting in significantly reduced re-accretion onto Mercury (\autoref{fig:coll_depletion}). Mercury accretes $\macc =8.2\times10^{-3}\,\Mearth$, only $8.2\%$ of the initial disk mass, compared to $m_{\rm acc,ND} = 4.3\times10^{-2}\,\Mearth$ without collisional depletion at the same time, implying that collisional depletion prevents approximately $81\%$ of the mass that would otherwise have been re-accreted from returning to Mercury. As the collision rate increases with disk mass, more massive disks are depleted more rapidly. The disk with $\Mdisk = 0.5\,\Mearth$ loses most of its mass by $1\,\mathrm{Myr}$, so that Mercury re-accretes only $\sim 3.3\%$ of the initial disk mass, although this corresponds to a larger absolute re-accreted mass, $\macc =1.64\times10^{-2}\,\Mearth$. In contrast, the least massive disk with $\Mdisk = 0.02\,\Mearth$ has a longer depletion timescale, and a larger fraction of the initial disk is re-accreted onto Mercury ($\sim 17.2\%$), although the absolute re-accreted mass remains small, $\macc=3.4\times10^{-3}\,\Mearth$.

To assess the impact of this reduced re-accretion on Mercury's composition, we compute the final CMF after re-accretion. We assume that proto-Mercury and the impactor initially had a terrestrial $\mathrm{CMF}_{\rm pre}$ of $\simeq 0.3$ before impact. We further assume that the giant impact strips much of proto-Mercury's mantle and leaves behind a remnant with Mercury's present-day mass, $M_{\rm p}=0.055\,\Mearth$, and ${\rm CMF}_{\rm p}=0.7$, before any subsequent re-accretion\footnote{A fully self-consistent treatment would require a separate immediate post-impact $M_{\rm p}$ and $\mathrm{CMF}_{\rm p}$ for each $\Mdisk$ to reproduce present-day Mercury after debris re-accretion.}.
Mass conservation of the core and mantle components gives the core mass fraction of the debris disk, ${\rm CMF}_{\rm d}=({\rm CMF}_{\rm pre}M_{\rm pre}-{\rm CMF}_{\rm p}M_{\rm p})/\Mdisk$, where $M_{\rm pre} (= M_{\rm p} + \Mdisk)$ is the combined pre-impact planet and impactor mass. For the canonical case with $\Mdisk=0.1\,\Mearth$, this gives a silicate-dominated disk with ${\rm CMF}_{\rm d}\simeq 0.08$. ${\rm CMF}_{\rm d}$ increases for more massive disks because the total initial core mass budget increases with $M_{\rm pre}$, while the core mass retained in the post-impact remnant is fixed by the assumed Mercury-like value, reaching ${\rm CMF}_{\rm d}\simeq 0.26$ for $\Mdisk=0.5\,\Mearth$. 
On the other hand, this approach gives an unphysical negative ${\rm CMF}_{\rm d}$ for very low-mass disks ($\Mdisk \lesssim 0.073\,\Mearth$). For such small $\Mdisk$, the adopted initial core mass fraction of $0.3$ implies that $M_{\rm pre}$ contains less core material than Mercury's present-day core. In such cases (e.g., $\Mdisk=0.02\,\Mearth$), we set ${\rm CMF}_{\rm d}=0$, corresponding to a purely silicate debris disk. Conserving the core mass then requires the combined proto-Mercury--impactor system to have had ${\rm CMF}_{\rm pre}\simeq0.51$. This may be interpreted as a multiple-impact scenario \citep[e.g.,][]{Chau_2018} in which Mercury's CMF was already enhanced by an earlier impact.

\begin{table*}
\centering
\caption{Pre-impact, post-impact, and post-accretion properties of Mercury for the three initial debris-disk masses considered, assuming large debris with $\Dmax=600\,\km$. All masses are in units of $\Mearth$. The subscript ``pre'' refers to the combined properties of the pre-impact planet and impactor, ``f'' refers to the final post-accretion values, and ``ND'' denotes values calculated without accounting for collisional depletion of the debris.}
\label{tab:cmf_estimates}
\small
\begin{tabular}{cccccccccccc}
\toprule
\multicolumn{2}{c}{\multirow{2}{*}{Pre-impact}} &
\multicolumn{10}{c}{Post-impact} \\
\cmidrule(lr){3-12}
\multicolumn{2}{c}{} & \multicolumn{4}{c}{Immediate} & \multicolumn{3}{c}{Post-accretion} & \multicolumn{3}{c}{Post-accretion, no debris depletion}\\
\cmidrule(lr){1-2}
\cmidrule(lr){3-6}
\cmidrule(lr){7-9}
\cmidrule(lr){10-12}
$M_{\rm pre}$ & $\mathrm{CMF}_{\rm pre}$ &
$\Mdisk$ & $\mathrm{CMF}_{\rm d}$ & $M_{\rm p}$ & $\mathrm{CMF}_{\rm p}$ &
$\macc$ & $M_{\rm p,f}$ & $\mathrm{CMF}_{\rm p,f}$ &
$m_{\rm acc,ND}$ & $M_{\rm p,f,ND}$ & $\mathrm{CMF}_{\rm p,f,ND}$\\
\midrule
0.075 & $0.51^{\dagger}$ &
0.02 & $0^{\dagger}$ & 0.055 & 0.70 &
0.003 & 0.058 & 0.66 &
0.009 & 0.064 & 0.60
\\
0.155 & 0.30 &
0.10 & 0.08 & 0.055 & 0.70 &
0.008 & 0.063 & 0.62 &
0.043 & 0.098 & 0.43
\\
0.555 & 0.30 &
0.50 & 0.26 & 0.055 & 0.70 &
0.016 & 0.071 & 0.60 &
0.217 & 0.272 & 0.35
\\
\bottomrule
\end{tabular}

\begin{flushleft}
\footnotesize{$^{\dagger}$ Adopting a terrestrial ${\rm CMF}_{\rm pre}=0.3$ would result in an unphysical negative ${\rm CMF}_{\rm d}$ for such a low mass disk. We therefore set ${\rm CMF}_{\rm d}=0$, corresponding to a purely silicate disk, and $\mathrm{CMF}_{\rm pre}=0.51$. This can be interpreted as a multiple-impact scenario in which Mercury's CMF had already been enhanced prior to the impact considered here.}
\end{flushleft}
\end{table*}

After re-accreting a mass $\macc$, we estimate the resulting CMF of Mercury as ${\rm CMF}_{\rm p,f}\simeq ({\rm CMF}_{\rm p}M_{\rm p}+{\rm CMF}_{\rm d}\macc)/(M_{\rm p}+\macc)$. The resulting values are summarized in \autoref{tab:cmf_estimates}. For the $\Mdisk = 0.1\,\Mearth$ case, with a total accreted mass of $\macc=8.2\times10^{-3}\,\Mearth$, we obtain a final CMF of $\simeq 0.62$, compared to $\simeq 0.43$ in the absence of collisional depletion. For the more massive disk, $\Mdisk = 0.5\,\Mearth$, the disk is more iron-rich (${\rm CMF}_{\rm d}\simeq 0.26$), and Mercury re-accretes $\macc=1.6\times10^{-2}\,\Mearth$ before faster collisional depletion removes much of the debris mass. The resulting final CMF is therefore $\simeq 0.60$, compared to $\simeq 0.35$ in the absence of collisional evolution. For the low-mass, purely silicate case with $\Mdisk = 0.02\,\Mearth$, the small absolute re-accreted mass gives a final CMF of $\sim 0.66$ at $10\,\mathrm{Myr}$, compared to $\sim 0.60$ without collisional depletion. Thus, when collisional depletion is accounted for, the final core mass fraction of Mercury does not fall below $\sim0.6$ even for the most massive disk, $\Mdisk = 0.5\,\Mearth$, with simulation parameters favoring re-accretion.

Overall, these results indicate that, for large, non-vaporized ejecta, collisional evolution and subsequent dust removal can significantly deplete the debris population and limit the re-accretion of mantle material onto Mercury, despite adopting parameters that favour re-accretion (e.g., large $\Dmax$ and small $\vinf$). In this regime, the disk remains optically thin, and radiation pressure efficiently removes micrometre-sized dust, truncating the collisional cascade and providing a pathway for mass loss that allows Mercury to retain a high core mass fraction.

% ------------------------------------------------------%
\subsection{Fate of vaporized debris}\label{subsec:vaporized_debris}
% ------------------------------------------------------%

\begin{figure}
\includegraphics[width=\columnwidth]{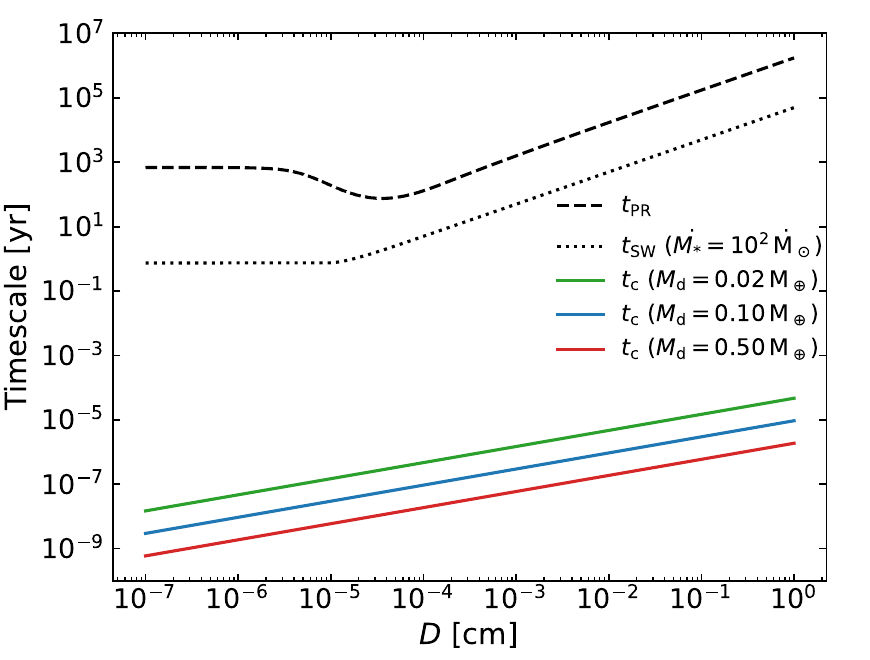}
\caption{
Poynting-Robertson drag (dashed), solar wind drag (dotted) and collisional (solid) timescales for small debris at $0.4\,\au$ as a function of particle diameter, $D$. The different coloured solid lines represent the collisional timescale for different initial disk masses, assuming $\Dmax = 1\,\mathrm{cm}$.
}
\label{fig:timescales}
\end{figure}

In this section, we explore the evolution of impact ejecta in the regime where the energy from a giant impact is sufficient to completely vaporize the material. The resulting vapor undergoes rapid expansion and cooling, ultimately condensing into millimetre to centimetre-sized droplets \citep{Melosh_Vickery_1991, Benz_2007, Johnson_Melosh_2014}. Particles in this size range, and smaller, are subject to significant non-gravitational forces in the inner Solar System, including Poynting–Robertson drag, solar wind drag, sputtering and electromagnetic forces, which can modify their orbital properties on timescales shorter than those associated with purely dynamical evolution \citep[e.g.,][]{Robertson_1937, Burns_1979, Mukai_Schwehm_1981, Gustafson_1994, Mann_2014, Baumann_2020}. Accordingly, we adopt an analytical approach to investigate their evolution in this regime, informed by the $N$-body simulations.
We first evaluate the efficiency of drag forces generally invoked in the literature for centimetre-sized condensates, while accounting for the disk’s optical thickness. We then explore the rapid collisional evolution of this debris and its implications for dust sizes, optical depth, and re-accretion. Finally, we discuss possible removal mechanisms responsible for clearing the dust grains produced in this collisional cascade from the inner Solar System.

% ------------------------------------------------------%
\subsubsection{Drag forces on centimetre-sized condensates}\label{subsubsec:drag}
% ------------------------------------------------------%

Poynting--Robertson (PR) drag continuously removes angular momentum from orbiting dust grains and drives inward migration on characteristic timescales \citep{Wyatt_Whipple_1950,Gladman_Coffey_2009}
\begin{equation}\label{eq:tpr}
    \tpr = \frac{1.68}{\Qpr} 
    \left(\frac{\rho}{3\,\gcc}\right)
    \left(\frac{D}{1\,\mathrm{cm}}\right)
    \left(\frac{a}{0.4\,\mathrm{au}}\right)^{2} \mathrm{Myr},
\end{equation}
where $\rho$, $D$, and $a$ are the particle density, diameter, and orbital radius, respectively, and $\Qpr$ is the PR drag efficiency factor. For centimetre-sized debris near Mercury's orbit ($a \sim 0.4\,\mathrm{au}$), assuming $\rho \sim 3\,\mathrm{g\,cm^{-3}}$ and $\Qpr \sim 1$, this yields $\tpr \approx 1.68\,\mathrm{Myr}$. This is comparable to the timescale over which a large fraction of debris is re-accreted onto Mercury (\autoref{fig:debris_accretion}), implying that PR drag alone is insufficient to efficiently remove ejecta before re-accretion \citep{Gladman_Coffey_2009}. 

Moreover, this estimate implicitly assumes that all particles are directly exposed to solar radiation, i.e., that the debris disk is optically thin. For the canonical impact, the cross-section of a $0.1\,\Mearth$ disk of centimetre-sized grains, $\Adust = 3\Mdisk/(2\rho D) \approx 3\times10^{26}\,\mathrm{cm^2}$, exceeds the disk area exposed to solar radiation, $\Adisk \sim 2\pi a\,\Delta h \approx 4\times10^{25}\,\mathrm{cm^2}$. The disk is therefore optically thick (for any $\Mdisk\gtrsim0.02\,\Mearth$). PR drag can only act efficiently on the exposed surface layers at the inner edge of the disk and would therefore take significantly longer than the single particle inspiral timescale presented in \autoref{eq:tpr} to remove the disk.

Solar-wind drag acts similarly to PR drag and contributes at a level $\sim1/3$ of the PR drag for the present-day Sun \citep{Burns_1979,Gustafson_1994}. The strength of winds in the young Sun, however, remains uncertain. While stronger winds with enhanced stellar mass-loss rates are often suggested to have been present during the period of terrestrial planet formation, with estimates of $\Mdot \sim 10$–$100\,\Mdotsolar$ \citep[e.g.,][]{Wood_2002, Johnstone_2015}, indirect observational constraints from young solar analogues suggest comparatively weak mass loss rates \citep{Wood_2014}. Adopting the enhanced wind scenario, \citet{Spalding_Adams_2020} showed that solar wind drag\footnote{In the enhanced wind regime with a rapidly rotating young Sun, the azimuthal component of the wind can become super-Keplerian, such that it transfers angular momentum to the grains and drives outward migration rather than inward drag.} can remove debris on timescales of
\footnote{The solar wind drag timescale used here differs from that quoted in \citet{Spalding_Adams_2020} because we adopt the convention of \citet{Wyatt_Whipple_1950}, in which the drag timescale represents the total inspiral time from a circular orbit at semimajor axis $a$ down to the Sun, whereas \citet{Spalding_Adams_2020} define a local orbital decay timescale as $a/|\dot{a}|$, which is larger than the inspiral time by a factor of $2$.}
\begin{equation}\label{eq:tsw}
\tsw = \frac{0.05}{\Qsw}
\left(\frac{\rho}{3\,\gcc}\right)
\left(\frac{D}{1\,\mathrm{cm}}\right)
\left(\frac{a}{0.4\,\mathrm{au}}\right)^{2}
\left(\frac{\Mdot}{100\,\Mdotsolar}\right)^{-1} \mathrm{Myr},
\end{equation}
where $\Qsw$ is the efficiency factor for momentum transfer of ions to debris particles. For centimetre-sized debris near Mercury's orbit \citep[$\Qsw \sim 1$,][]{Minato_2004}, this corresponds to sub-Myr drift timescales for $\Mdot \sim 100\,\Mdotsolar$, shorter than both re-accretion and PR drag timescales. However, if solar winds in young Sun were weak, the efficiency of this mechanism would be correspondingly reduced.

% ------------------------------------------------------%
\subsubsection{Collisional evolution}\label{subsubsec:coll_evol}
% ------------------------------------------------------%

However, the estimates of removal timescales through drag forces made above neglect the collisional evolution of the debris. In a dense debris disk at sub-au orbital radii, centimetre-sized bodies undergo frequent mutual collisions \citep{Benz_1988}, leading to a collisional cascade that grinds material to progressively smaller sizes. Within such a cascade, the collisional lifetime of bodies of different sizes can be estimated from the collisional mass-loss rate of \citet[Eq.~A20]{Ghosh_2025} as
\begin{align}\label{eq:tcol}
\tcol \simeq\;&
9.4\times10^{-12}
\left(\frac{\rho}{3\,\mathrm{g\,cm^{-3}}}\right)
\left(\frac{a}{0.4\,\mathrm{au}}\right)
\left(\frac{\Delta a}{0.25\,\mathrm{au}}\right)
\left(\frac{\Delta h}{0.074\,\mathrm{au}}\right)\nonumber\\
& \hspace{1.5cm} \times
\left(\frac{\frho}{0.02}\right)
\left(\frac{\vrel}{2.5\,\mathrm{km\,s^{-1}}}\right)^{-8/3}
\left(\frac{\Qdstar}{5\times10^3\,\mathrm{J\,kg^{-1}}}\right)^{5/6}\nonumber\\
& \hspace{1.5cm} \times
\left(\frac{\Mdisk}{0.1\,\Mearth}\right)^{-1}
\left(\frac{\Dmax}{1\,\mathrm{cm}}\right)^{1/2}
\left(\frac{D}{1\,\mathrm{cm}}\right)^{1/2}
\ \mathrm{Myr}.
\end{align}
The estimate assumes an effective disk volume $V_{\rm eff}\approx \frho 2\pi a\Delta a\Delta h$. Here, $\Delta a$ and $\Delta h$ are the $99^\mathrm{th}$ percentile radial and vertical extents of the debris disk, respectively, while $\frho$ accounts for the fact that most of the debris occupies a more spatially concentrated region within these extents. These percentile-defined extents provide near-maximal measures of the radial width and vertical thickness of the debris distribution and therefore overestimate the volume occupied by most collisionally active debris. Moreover, \autoref{eq:tcol} assumes that the debris uniformly fills an axisymmetric disk, whereas the numerical collisional evolution in \autoref{subsec:large_debris} uses the three-dimensional debris distribution obtained from the $N$-body simulations. We therefore treat $\frho$ as an empirical correction factor and calibrate it by requiring the analytical and numerical collision timescales to agree. We use the most massive ($\Mdisk=0.5\,\Mearth$) large-fragment disk ($D_{\max}=600\,\km$) for this calibration, as this disk collisionally depletes earliest, when the debris is still spatially concentrated, with a pronounced density enhancement near the collision point that produces higher local collision rates relative to the rest of the disk \citep[e.g.,][]{Jackson_2014, Jones_2023}. We find that setting $\frho=0.02$ ensures \autoref{eq:tcol}, evaluated at $D=D_{\max}=600\,\km$ with $\Qdstar\simeq1.2\times10^6\,\Jkg$ for $\Mdisk=0.5\,\Mearth$, gives $\tcol\sim10^4\,\mathrm{yr}$, consistent with the numerical depletion timescale found in \autoref{subsec:large_debris} and shown in \autoref{fig:coll_depletion}. The value $\frho<1$ therefore represents the smaller effective volume, and correspondingly higher density, of the collisionally active debris relative to a uniform axisymmetric disk filling the full $99^\mathrm{th}$ percentile-defined volume.

The extremely short collisional lifetimes inferred for vapor condensates therefore follow from applying this numerically anchored scaling to much smaller particles. For mm-cm-sized condensates, the same disk mass is distributed among many more particles than in the large-fragment case, producing a much higher number density and rate of mutual collisions. Hence, once destructive relative velocities are established, \autoref{eq:tcol} implies that the collisional cascade proceeds rapidly. However, \autoref{eq:tcol} does not account for the preceding evolution during which the ejecta initially move away from one another before their heliocentric orbits bring them back into orbit-crossing configurations. $\tcol$ should therefore be interpreted as the collisional timescale after the ejecta form a mutually colliding population, rather than as the lifetime of the debris since the primary impact.

\autoref{fig:timescales} compares the collisional timescales with Poynting–Robertson drag and solar wind drag timescales as a function of particle size. We find that the collisional lifetimes are several orders of magnitude shorter than the corresponding drag timescales. As a result, particles undergo rapid collisions before drag forces can significantly alter their orbits \citep[e.g.,][]{Wyatt_2005}. In optically thin disks, such a collisional cascade extends down to the radiation pressure blowout limit (as discussed in \autoref{subsec:large_debris}), below which particles are efficiently removed by radiation pressure \citep[e.g.,][]{Wyatt_2008}.

In the optically thick debris cloud considered here, this picture no longer holds. Because the total cross-sectional area is dominated by the smallest grains, the collisional cascade leads to a rapid increase in the total dust cross-section as the mm-cm-sized condensates are collisionally ground down into progressively smaller particles on sub-orbital timescales. For a steady-state collisional cascade, the geometric cross-section scales as $\Adustgeom \sim \Dmin^{-1/2}$ (\autoref{eq:adust}), implying that the optical depth increases as the cascade proceeds to smaller sizes. As a result, the disk becomes increasingly optically thick, and the fraction of debris particles exposed to solar radiation becomes progressively smaller, reducing the efficiency of radiation pressure in removing the grains. Consequently, particles are not efficiently removed at the nominal radiation pressure blowout limit ($\Dbl$), and the collisional cascade can continue to smaller sizes.

\begin{figure}
\includegraphics[width=\columnwidth]{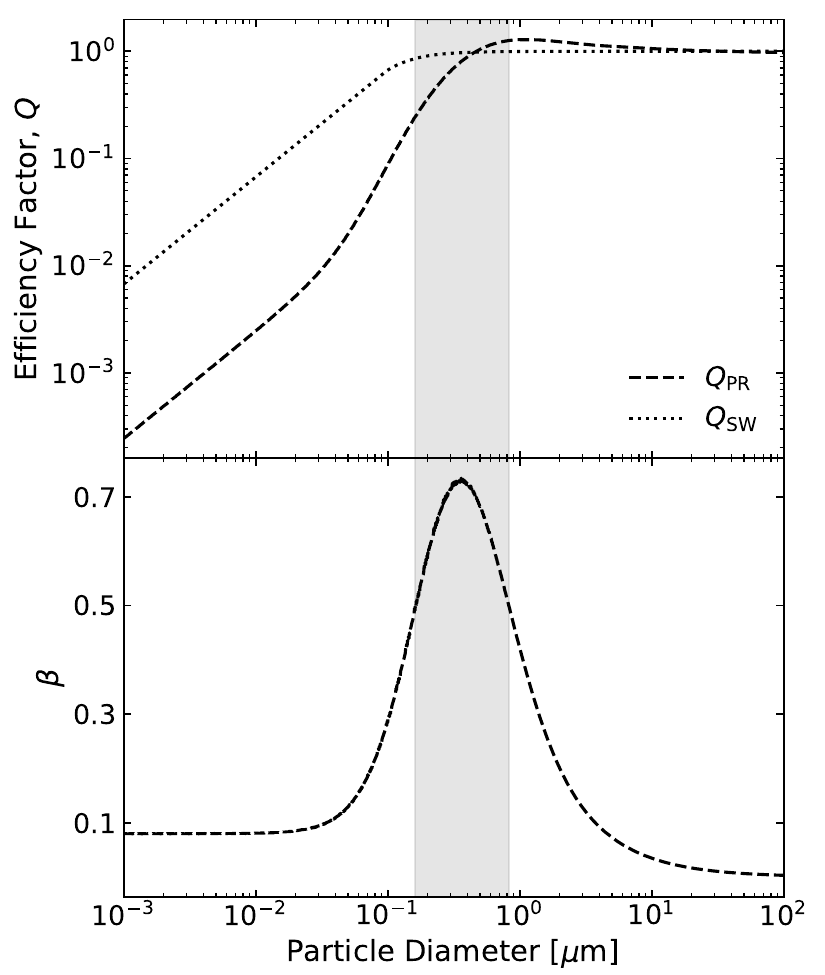}
\caption{
\textit{Top}: Radiation-pressure efficiency, $\Qpr$ (dashed curve), and solar wind momentum-coupling efficiency, $\Qsw$ (dotted curve), as a function of particle diameter, $D$. 
\textit{Bottom}: Ratio of radiation pressure to gravity, $\beta$. 
The grey shaded region marks the radiation-pressure blowout regime ($\beta > 0.5$).
}
\label{fig:dust_rad_props}
\end{figure}
\begin{figure}
\includegraphics[width=\columnwidth]{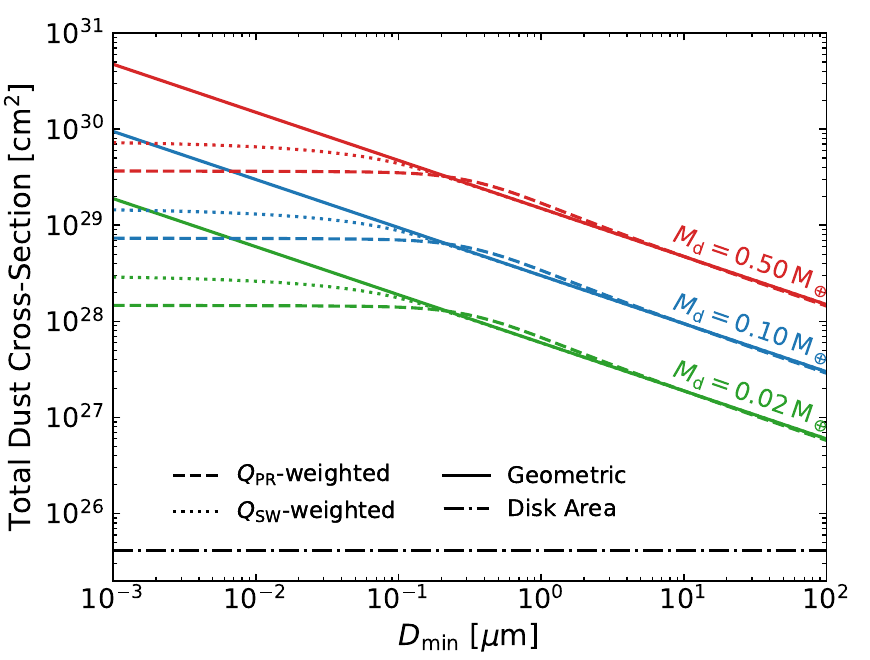}
\caption{
Total cross-sectional area of particles in a steady-state collisional cascade as a function of the minimum particle size, $\Dmin$, assuming $\Dmax = 1\,\mathrm{cm}$. The solid curves show the geometric cross-section, $\Adustgeom$ (\autoref{eq:adust}), while the dashed and dotted curves show the effective cross-sections weighted by the radiation-pressure efficiency ($\Qpr$), and the solar wind coupling efficiency ($\Qsw$), respectively. The dash-dotted line indicates the area of the disk exposed to solar radiation and wind, $\Adisk \sim 2\pi a\,\Delta h$. Different colours correspond to different disk masses.
}
\label{fig:sigma_dmin}
\end{figure}
In this regime, as well as optical depth effects, it is also important to account for the fact that small grains interact less efficiently with radiation below sizes comparable to the peak wavelength of the solar spectrum, so that the assumption that $\Qpr\sim 1$ is no longer valid. To take this into account, we calculate the radiation pressure efficiency factor, $\Qpr$, as a function of particle size for the solar spectrum using Mie theory with the {\tt astromiedust}\footnote{\url{https://github.com/rigilkent/astromiedust}} package \citep{Sommer_2025}. We adopt compact, non-porous silicate grains with no water-ice component (\texttt{qsil}=1, \texttt{qice}=0, and \texttt{mpor}=0). \autoref{fig:dust_rad_props} shows the size-dependent coupling of silicate grains to solar radiation. The top panel shows that $\Qpr \approx 1$ for particles larger than a few microns, but declines sharply at smaller sizes as grains become inefficient absorbers of solar radiation. The corresponding ratio of radiation pressure to gravity, $\beta$, is shown in the bottom panel. As particle size decreases, the individual particle mass declines faster than the individual cross-section, leading to an increase in $\beta$, which exceeds the blowout threshold $\beta \gtrsim 0.5$ near $D \sim 0.8\,\um$ and peaks at $\beta \sim 0.73$ around $D \sim 0.36\,\um$. At smaller sizes, $\beta$ decreases as $\Qpr$ declines, dropping below $0.5$ at $D \sim 0.16\,\um$ and saturating at $\sim 0.08$ for $D \lesssim 0.02\,\um$. Thus, below $D \sim 0.16\,\um$, gravity again dominates over radiation pressure, preventing radiation pressure from unbinding and removing the exposed grains.\footnote{Note that the exact values of $\beta$ depend on the dust composition, although the overall shape of the $\beta$–$D$ curve is expected to be similar for all compositions.}

Since small grains interact less efficiently with radiation, their effective cross-section is reduced accordingly. As the collisional cascade produces smaller particles and lowers $\Dmin$, the geometric cross-section continues to increase with decreasing $\Dmin$ (\autoref{eq:adust}), but the $\Qpr$-weighted effective cross-section, $\Adustpr$, eventually saturates as $\Qpr$ declines for small particles (\autoref{fig:sigma_dmin}).
Despite this saturation, $\Adustpr$ remains several orders of magnitude larger than $\Adisk$, resulting in only a very small fraction of particles being directly exposed to solar radiation, $\fexppr \sim \Adisk/\Adustpr \sim 5.7\times 10^{-4}(\Mdisk/0.1\,\Mearth)^{-1}$. Thus, only a small fraction of the debris is subject to radiation-driven removal. Radiation pressure therefore cannot effectively truncate the collisional cascade. Instead, the evolution of the debris is governed by collisional processes.

In the absence of efficient removal, these grains continue to undergo frequent mutual collisions, driving a collisional cascade toward progressively smaller sizes. However, another effect to consider is that at sufficiently small sizes, the available kinetic energy in collisions becomes insufficient to create new surface area, and collisions transition from being fragmenting to non-destructive \citep{Krijt_Kama_2014}. This transition occurs at a size,
\begin{equation}
\Dminlim \simeq 2.7
\left(\frac{\gamma}{0.05 \,\mathrm{J\,m^{-2}}}\right)
\left(\frac{\eta}{0.01}\right)^{-1}
\left(\frac{\rho}{3 \,\mathrm{g\,cm^{-3}}}\right)^{-1}
\left(\frac{\vrel}{2.5 \,\mathrm{km\,s^{-1}}}\right)^{-2}
\mathrm{nm}.
\end{equation}
Here, $\gamma$ is the surface energy per unit area and $\eta$ is the fraction of the impact energy used to create new surface. For our nominal relative velocity, this corresponds to $\Dminlim \simeq 2.7\,\mathrm{nm}$, placing the characteristic fragmentation limit in the nanometre-size regime, subject to uncertainties in $\gamma$ and $\eta$. The existence of dust particles at these sizes is observationally inferred from spacecraft observations in the Solar System, including dust streams from the Jovian and Saturnian systems and fast nanodust inferred in the inner heliosphere \citep[e.g.,][]{Mann_2014,Poppe_Lee_2020}.

The disruption threshold for nanometre-sized silicate grains is predicted to be $\Qdstar \simeq 2\times10^6\,\Jkg$ \citep{Benz_Asphaug_1999}, while the specific impact energy of equal-sized collisions at $\vrel=2.5\,\kms$ is $Q_{\rm col}=\vrel^2/2\simeq3\times10^6\,\Jkg$. Although $\Qdstar$ remains highly uncertain at these sizes \citep[e.g.,][]{Stein_2026}, these estimates also place equal-sized collisions near the disruption threshold, suggesting that further fragmentation becomes inefficient in the nanometre regime.

Notably, $\Dminlim$ is derived for equal-sized, fully destructive collisions. Particles of this size can still be destroyed by impacts with larger bodies, and smaller grains may still be produced in energetically allowed collisions in different regime of the cascade. Hence, $\Dminlim$ does not represent a sharp lower cutoff of the size distribution. Numerical studies that incorporate this surface-energy constraint into collisional cascade models show that its global effect is modest, because a substantial fraction of dust is produced in high-velocity collisions that are not limited by this constraint \citep{Thebault_2016}. Instead, $\Dminlim$ marks the scale below which fragmentation becomes increasingly inefficient in mutual collisions. Overall this indicates that in the optically thick debris cloud at Mercury’s orbital distance the collisional cascade can extend well below the radiation pressure blowout limit and extend to nanometre scales.

% ------------------------------------------------------%
\subsubsection{Collisional damping and nanodust re-accretion}\label{subsubsec:damping_re-accretion}
% ------------------------------------------------------%
Near $\Dminlim$, the collisional cascade is likely to extend into a regime where collisions become less efficient at catastrophic fragmentation and increasingly dissipative. However, as long as these grains remain embedded in a dynamically hotter disk undergoing fragmenting collisions, they are continuously stirred and subjected to destructive impacts by larger fragments. Nevertheless, this phase is expected to be short-lived given the extremely short collisional lifetimes of the larger grains (\autoref{eq:tcol}, also see \autoref{fig:timescales}).

\begin{figure*}
    \centering
    \includegraphics[width=0.4\textwidth]{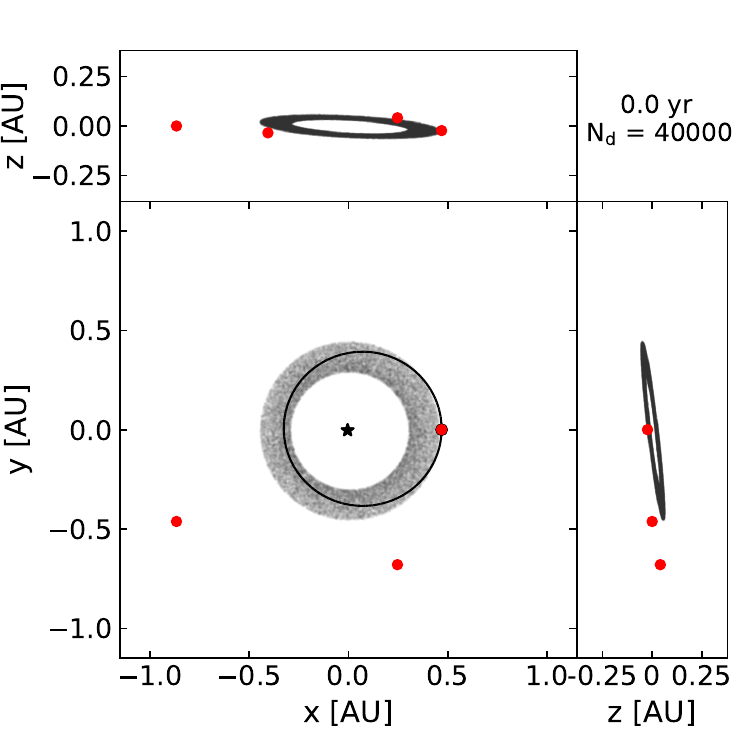}\hspace{1em}
    \includegraphics[width=0.525\textwidth]{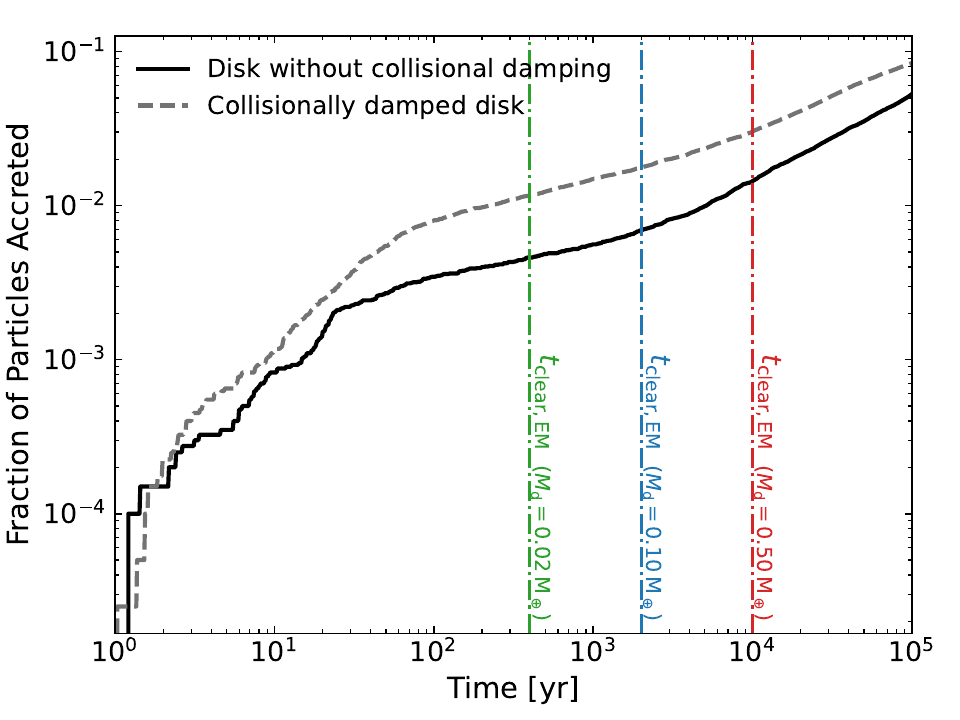}\hspace{0.1em}
    \vspace{1ex}
    \caption{
    \textit{Left panel}: Initial disk snapshot for debris particles initialized on equivalent circular orbits (\autoref{subsubsec:damping_re-accretion}), shown with the same axes and layout as in \autoref{fig:snapshots}. Mercury’s orbit is shown in black in the $x$–$y$ plane for reference.
    \textit{Right panel}: Cumulative fraction of debris particles accreted onto Mercury over time for the collisionally damped, nearly circular and coplanar disk (gray dashed line), compared with the disk without collisional damping from the simulations presented in \autoref{sec:res} (black solid line, also see \autoref{fig:debris_accretion}). The vertical dash-dotted lines show the estimated disk-clearing timescales due to electromagnetic (EM) removal for different disk masses (\autoref{eq:temclr}).
    }
    \label{fig:thin_disk}
\end{figure*}

Once the larger grains have been collisionally depleted, most of the mass is expected to reside in nanometre-sized dust undergoing frequent non-fragmenting, dissipative collisions. In this regime, repeated collisions between nanodust grains dissipate random kinetic energy, damping their eccentricities and inclinations. This damping would drive a collapse of the nanodust population toward circular coplanar orbits in the disk mid plane, producing a geometrically thin annulus.
Once its vertical thickness becomes comparable to or smaller than the Sun's projected radius, the top and bottom surfaces of the annulus can be directly exposed to solar radiation and wind. The projected exposed surface area is then $\Athindisk \sim 2 \times 2\pi a \Delta a^{\prime} \times (R_{\sun}/a) \sim 4 \pi \delaprime R_{\sun}$, where primes denote quantities of the collisionally damped disk of nanodust.
We estimate the radial width of this circularized annulus, $\delaprime$, from the distribution of equivalent circular semi-major axes of the debris shortly after ejection. These are defined as the semi-major axes of circular orbits with the same vertical angular momentum as the original debris orbits \citep[e.g.,][]{Canup_2004},
\begin{equation}\label{eq:acirc}
    a_{\rm eq} = a(1-e^2)\cos^2 i,
\end{equation}
where $i$ is the inclination with respect to the disk midplane. The equivalent circularized orbits of the ejected debris particles, shown in the left panel of \autoref{fig:thin_disk}, have a radial extent of $\Delta a^{\prime} \sim 0.15\,\au$ \footnote{Depending on the exact location of the impact (e.g. near pericentre or apocentre), the ejected debris may occupy slightly different orbits \citep{Jackson_2014}, leading to small variations in our estimate of $\delaprime$. However, such small differences would have a minimal impact on our results.}. This makes $\Athindisk$ $\sim 21$ times smaller than $\Adisk$, reducing the fraction of the optically thick dust surface directly exposed to solar radiation and wind by the same factor.

As the nanodust disk becomes geometrically thin, Keplerian shear generates random motions through viscous stirring, while inelastic collisions dissipate them \citep[e.g.,][]{Brahic_1977, Goldreich_Tremaine_1978}. In the regime where the grains undergo multiple collisions per orbit, the nanodust population can be treated as a viscous fluid, with a characteristic radial spreading timescale, $\tvisc \sim \delaprime^2/(\sigma^2 \tcol)$, where $\sigma$ is the velocity dispersion and $\tcol$ is the non-fragmenting collision time of the nanodust \citep[][]{Goldreich_Tremaine_1978}. We estimate $\sigma$ when the vertical thickness becomes comparable to the Sun's projected radius, giving $\sigma \sim (R_{\sun}/a)\vk$.\footnote{The precise vertical thickness of the nanodust disk is uncertain, hence we evaluate $\sigma$ at the characteristic scale set by the Sun's projected radius. If the annulus damps to a smaller thickness, $\sigma$ decreases and $\tvisc$ increases.} The collision time can be estimated as $\tcol \sim \tper/(4\pi\taueff)$, where $\tper$ is the orbital period, $\taueff = \Ananodustgeom/(2\pi a\delaprime)$ is the face-on geometric optical depth of the nanodust annulus, and $\Ananodustgeom = 3\Mdisk/(2\rho\Dminlim)$ is the total geometric cross-sectional area of the nanodust population \citep{Wyatt_1999}. Thus,
\begin{align}\label{eq:tvisc}
\tvisc \gtrsim\;&
1.1\times 10^9\,\mathrm{yr}
\left(\frac{\Mdisk}{0.1\,\Mearth}\right)
\left(\frac{\rho}{3\,\gcc}\right)^{-1}
\left(\frac{\Dminlim}{2.7\,\mathrm{nm}}\right)^{-1}\nonumber\\
& \hspace{1.25cm} \times
\left(\frac{a}{0.4\,\au}\right)^{1/2}
\left(\frac{\delaprime}{0.15\,\au}\right).
\end{align}
The inequality reflects that any further reduction in the vertical thickness below the scale set by the Sun's angular size would reduce $\sigma$ and therefore increase $\tvisc$. The extremely long viscous timescale suggests that viscous spreading is unlikely to substantially broaden the disk before the dust evolves through other processes.

The evolution and accretion of the circularized nanodust in this collisionally damped, geometrically thin disk will differ from that in the fiducial simulations presented at the beginning of \autoref{sec:res}, where the debris forms a dynamically hot disk with a significant vertical width (\autoref{fig:snapshots}). To estimate the amount of re-accretion from the damped disk, we conduct an additional idealized $N$-body simulation in which the debris is initialized in Mercury's orbital plane, with semimajor axes set to the equivalent circular semimajor axes from \autoref{eq:acirc} and negligible eccentricities and inclinations ($\sim10^{-5}$). All other simulation parameters are kept unchanged. The resulting initial configuration is shown in the left panel of \autoref{fig:thin_disk}. We integrate the system for $10^5\,\mathrm{yr}$ and track the cumulative fraction of debris particles re-accreted onto the planets.

The right panel of \autoref{fig:thin_disk} shows that Mercury re-accretes debris more rapidly from a collisionally damped, nearly circular and coplanar disk than from the disk without collisional damping. The enhancement becomes significant after $\sim 10\,\mathrm{yr}$ and reaches a maximum factor of $\simeq 2.8$ at $t\simeq 1.9\times10^2\,\mathrm{yr}$. At this time, the re-accreted fraction is $\simeq 10^{-2}$ in the damped disk, compared to $\simeq 3.5\times10^{-3}$ in the disk without damping. The enhancement remains comparable at $10^3\,\mathrm{yr}$, with re-accreted fractions of $\simeq 1.5\times10^{-2}$ and $\simeq 5.6\times10^{-3}$, respectively. At later times, planetary perturbations increase the radial and vertical extent of the disk, and the difference between the two cases decreases. By $10^5\,\mathrm{yr}$, the re-accreted fraction reaches $\simeq 8.4\%$ in the collisionally damped disk, compared to $\simeq 5.6\%$ in the disk without damping. No other planet accretes any particles over the duration of the simulation.
However, this purely dynamical calculation does not include the rapid dissipative collisions expected for nanometre-sized grains. Any orbital excitation produced by planetary perturbations would therefore likely be damped on a short collisional timescale. Conversely, particles perturbed out of the disk plane before undergoing subsequent collisions would become directly exposed to solar radiation and wind, and may be rapidly removed. These effects are not captured in our $N$-body simulation. Nonetheless, as we show in the next subsection, the nanodust removal timescales are much shorter than the timescale on which planetary perturbations introduce a significant disk height. The simplified $N$-body model and short integration duration are therefore sufficient for estimating dust re-accretion in this regime.

% ------------------------------------------------------%
\subsubsection{Removal of nano-dust from optically thick disk}\label{subsubsec:dust_removal}
% ------------------------------------------------------%

We now consider the possible removal mechanisms acting on nanodust grains in a geometrically thin, optically thick disk at Mercury’s orbital distance.

\textbf{PR drag:}
At nanometer sizes, $\Qpr$ decreases almost linearly with decreasing particle size (\autoref{fig:dust_rad_props}), such that the PR drag timescale (\autoref{eq:tpr}) for the exposed dust approaches a constant value of $\tpr \approx 689\,$yr (\autoref{fig:timescales}). Once the debris have collisionally evolved into nanodust grains of size $\Dminlim$, only a very small fraction $\fexpprthin \sim \Athindisk/\Adustprthin \sim 2.7\times10^{-6}(\Mdisk/0.1\,\Mearth)^{-1}$ of the dust is exposed to solar radiation at any given time, where $\Adustprthin$ is the total effective cross-sectional area of this single-size nanodust population, $\Adustprthin \approx \Qpr(\Dminlim) \times 3\Mdisk/(2\rho\Dminlim)$. 
The global disk-clearing time ($\tclrpr$) can therefore be estimated by treating the exposed nanodust fraction as the portion of the disk actively losing angular momentum to PR drag. The corresponding mass-loss rate is $\Mdiskdotpr \sim \fexpprthin \Mdisk/\tpr$, giving
\begin{equation}
    \tclrpr \sim \frac{\Mdisk}{\Mdiskdotpr} \sim \frac{\tpr}{\fexpprthin}
    \sim 2.5\times10^{8}
    \left(\frac{\Mdisk}{0.1\,\Mearth}\right)\,\mathrm{yr}.
    \label{eq:tclr_pr}
\end{equation}
As this timescale is substantially longer than the re-accretion timescale, PR drag alone is unlikely to have any significant impact on debris re-accretion.

\textbf{Solar wind drag:}
In a similar manner, the solar wind coupling efficiency, $\Qsw$, decreases for small grains, as the solar wind particles only partially deposit their momentum when interacting with the grain \citep{Minato_2004}. In contrast, for larger grains $\Qsw \sim 1$, as the incident solar wind particles are effectively stopped and transfer their full momentum. Adopting a solar wind composed of protons and $\alpha$-particles in the ratio $1:0.05$, and neglecting heavier species, we compute $\Qsw$ as a function of grain size (top panel of \autoref{fig:dust_rad_props}) for SiO$_2$ following \citet{Minato_2004}. We find that the solar wind drag timescale saturates at $\tsw \sim 0.75\,\mathrm{yr}$ for $D \lesssim100\,$nm, assuming $\Mdot \sim 100\,\Mdotsolar$ (\autoref{fig:timescales}) at Mercury's orbital radius. However, only a fraction of the dust is directly exposed to the solar wind, given by $\fexpswthin \sim \Athindisk/\Adustswthin \sim 9.8\times10^{-8}(\Mdisk/0.1\,\Mearth)^{-1}$, where $\Adustswthin = \Qsw(\Dminlim) \times 3\Mdisk/(2\rho \Dminlim)$ is the $\Qsw$-weighted total cross-section of the nanodust population. Thus, the overall disk clearing time can be estimated as
\begin{equation}
    \tclrsw \sim \frac{\tsw}{\fexpswthin}
    \sim 7.7\times10^{6}
    \left(\frac{\Mdisk}{0.1\,\Mearth}\right)\,\mathrm{yr}.
    \label{eq:tclr_sw}
\end{equation}
Thus, even in the enhanced-wind limit ($\Mdot = 100\times\Mdotsolar$), the disk-clearing timescale due to solar-wind drag remains long enough to allow substantial re-accretion of the debris.

\textbf{Lorentz force:}
At nanometre scales, dust grains are expected to become charged and strongly coupled to the solar wind plasma. Photoelectric emission, whereby incident UV photons eject electrons from the grain surface, together with interactions with the ambient plasma, leads to rapid charging of small grains \citep[e.g.,][]{Mann_2014}. The resulting large charge-to-mass ratios imply that electromagnetic forces can become comparable to, or exceed, gravity, such that grain dynamics are governed primarily by the Lorentz force associated with the magnetic field carried by the solar wind. In this regime, charged nanograins behave analogously to pickup particles, being accelerated by the electromagnetic field to velocities of order hundreds of $\kms$ \citep{Czechowski_Mann_2010, Mann_2014, Poppe_Lee_2020}. This provides an efficient removal mechanism, with a characteristic transport timescale $\tem \sim \delaprime / v_{\rm sw} \approx 0.9\,\mathrm{day}$, with $v_{\rm sw} \sim 300\,\kms$.

However, in an optically thick debris cloud, this process is limited by the rate at which the exposed surface layers can be charged. Once charged, these grains are removed on the short electromagnetic transport timescale, which then exposes a new layer to charging. The global removal rate is therefore set primarily by the charging rate from the ionizing solar UV flux. In contrast, PR and solar-wind drag act much more slowly on the exposed grains, so grain removal is set by the small exposed surface area of the disk. For Lorentz force removal, the charging rate can then be estimated from the photoelectron flux from an uncharged grain, which is set by the ionizing solar photon flux above the photoelectric threshold, weighted by the photoelectric yield and the photon absorption cross-section. The resulting photoelectron emission rate per unit grain surface area under the present-day solar UV flux
\footnote{Mercury's giant impact would have occurred in the young Solar System, when the solar spectrum and luminosity may have differed from present-day values \citep[e.g.,][]{Ribas_2005}. We therefore use the present-day solar flux as a reference normalization.}
can be estimated as \citep[][]{Mann_2014}
\begin{equation}
R_{\rm ph0} \simeq 0.5\times10^{10}\chi \left(\frac{a}{\au}\right)^{-2}\,\mathrm{cm^{-2}\,s^{-1}},
\end{equation}
where $\chi$ is a dimensionless efficiency factor, with $\chi\sim1$ for silicates. The total electron production rate over the exposed surface of the disk is then $\dot N_e = R_{\rm ph0} \Athindisk$. For efficient electromagnetic removal, grains must reach a charge-to-mass ratio of at least $Q/m \sim 10^{-5}\,e/\mproton$ \citep{Czechowski_Mann_2010}, corresponding to $\sim 10\,$nm grains. The maximum amount of disk mass that can be charged to this threshold per unit time, through the exposed surface layer, is therefore
\begin{equation}
\dot M_{\rm ch} \sim \frac{e\dot N_e}{Q/m} \simeq \frac{R_{\rm ph0}\Athindisk \mproton}{10^{-5}}.
\end{equation}
The corresponding disk clearing timescale can then be estimated as
\begin{equation}\label{eq:temclr}
\tclrem \sim \frac{\Mdisk}{\dot M_{\rm ch}}
\sim \frac{2\times10^3}{\chi}\,
\left(\frac{\Mdisk}{0.1\,\Mearth}\right)
\left(\frac{a}{0.4\,\au}\right)^{2}
\left(\frac{\delaprime}{0.15\,\au}\right)^{-1} \mathrm{yr}.
\end{equation}
For the canonical disk mass ($\Mdisk=0.1\,\Mearth$) and $\chi\sim1$, this gives 
$\tclrem \sim 2\times10^3\,\mathrm{yr}$. Over this timescale, the purely dynamical simulation suggests that Mercury would re-accrete only $\simeq 1.8\%$ of the debris from the collisionally damped disk (\autoref{fig:thin_disk}, right panel). Thus, electromagnetic removal can rapidly deplete the exposed nanodust layers before substantial dynamical re-accretion occurs. Since this is the shortest disk-clearing timescale among the removal processes considered here, electromagnetic removal is likely to be the dominant pathway for removing nanodust from the optically thick, geometrically thin disk.

The estimate of $\tclrem$ above is subject to significant uncertainties. On the one hand, it assumes the present-day solar flux, whereas the young Sun is expected to have had a substantially stronger high-energy photoionizing flux \citep[e.g.,][]{Ribas_2005}. On the other hand, it assumes efficient photoelectric charging, neglects the suppression of photoelectron emission once grains become positively charged, and assumes that charged nanograins are removed rapidly enough to continuously expose a fresh surface layer.
The corresponding mass-loss rate, and hence $\tclrem$, should therefore be treated as an order-of-magnitude estimate, with significant uncertainties in both directions. The use of the present-day solar flux may underestimate the charging rate and thus overestimate $\tclrem$, whereas the assumed efficiency of photoelectric charging and charged-grain removal may overestimate the effective removal rate and thus underestimate $\tclrem$.
The fate of the charged grains also depends on their subsequent dynamical evolution, which determines whether they escape directly or undergo further interactions with disk particles.

A full treatment of the dynamical trajectories of charged nanograins, including the Lorentz force, is beyond the scope of this work. Previous studies show that charged nanograins are not simply accelerated along purely radial escape trajectories; Lorentz forces can also drive latitudinal, i.e. out-of-plane, motion associated with the heliospheric magnetic field and current sheet \citep[e.g.,][]{Poppe_Lee_2020}. Thus, while some charged nanograins may be accelerated out of the disk plane and escape directly, others may retain a significant in-plane component and collide with particles near the exposed surface of the optically thick disk. At the velocities expected from electromagnetic pickup (potentially hundreds of $\kms$), these impacts are highly energetic, are likely to be destructive, and can lead to vaporization \citep[e.g.,][]{Hornung_2000}. Since these collisions are expected to occur in the exposed layers, the resulting vapor remains subject to solar UV irradiation and the ambient plasma environment, allowing it to become ionized. The resulting ions have large charge-to-mass ratios, are strongly coupled to the solar magnetic field, and can be removed as pickup ions. Impacts of these accelerated grains with the disk may also partially offset collisional damping and help maintain a finite vertical thickness, exposing the radial inner edge to solar radiation and the solar wind in addition to the upper and lower surfaces. Thus, even when charged nanograins do not escape directly, electromagnetic acceleration may still provide a fast mass loss pathway through destructive surface-layer collisions followed by ion pickup. Among the removal processes considered here, electromagnetic removal gives the shortest disk clearing timescale for Mercury's giant impact debris disk, potentially as short as a few thousand years once the collisional cascade in optically thick disks produces nanometre-sized grains that become charged by solar UV radiation.

\textbf{Sputtering:}
Solar wind sputtering provides an additional erosion mechanism for nanometre-sized grains through the removal of surface atoms by energetic ion impacts. 
While individual nanograins can be eroded on short timescales ($\tsput \sim \mathrm{few}\times100\,$days) at Mercury's orbital distance under present-day solar-wind conditions \citep{Baumann_2020}, the total sputtering-driven mass-loss rate from the disk is also limited by the surface area directly exposed to the solar wind. 
Following \citet{Baumann_2020}, we estimate
\begin{equation}
\mdotsput = \nsw \vsw \Athindisk \Ytot \matom,
\end{equation}
where $\nsw$ and $\vsw$ are the solar wind ion density and velocity, $\Ytot\sim 0.01$ is the sputtering yield, $\matom$ is the mean mass of sputtered atoms. Adopting solar wind parameters from \citet{Baumann_2020} scaled to $a = 0.4\,\au$, we obtain $\mdotsput \sim 10^{9}\,\mathrm{g\,s^{-1}}$. For a disk mass $\Mdisk=0.1\,\Mearth$, this corresponds to a clearing time $\tclrsput \sim \Mdisk/\mdotsput \gtrsim 10^{10}\,\mathrm{yr}$. For an enhanced solar wind with $\Mdot \sim 100\,\Mdotsolar$ \citep{Spalding_Adams_2020}, the mass-loss rate would increase proportionally, reducing the corresponding timescale by a similar factor. Thus, although sputtering can efficiently erode individual nanograins, it is insufficient to remove the bulk disk mass on sub-Myr timescales.

\textbf{Mass dependence:}
For all of the removal mechanisms considered above, the dust removal rate in the optically thick regime is limited by the exposed geometric area of the disk. For a fixed disk geometry, this exposed area is largely independent of $\Mdisk$. Since a larger disk mass must be removed through the same exposed surface area, the clearing time increases linearly with mass, $\tclr \propto \Mdisk$, provided that the disk remains optically thick and that the collisional cascade supplies small grains faster than they are removed. More massive disks are therefore depleted proportionally more slowly, while lower-mass optically thick disks are cleared faster. This behaviour differs from the optically thin regime discussed in \autoref{subsec:large_debris}, where the mass-loss rate is controlled by the rate at which the collisional cascade supplies grains down to $\Dbl$. In that case, the collision rate increases with disk mass, so more massive disks deplete faster (\autoref{fig:coll_depletion}).

% ------------------------------------------------------%
\textbf{Summary:}
Overall, these results show that in optically thick, massive debris disks at Mercury's orbital distance, collisional evolution can rapidly produce nanometre-sized grains that become charged and are subsequently removed by electromagnetic forces on timescales as short as a few thousand years. This is more than an order of magnitude shorter than in the regime where the debris is dominated by large, unvaporized fragments (\autoref{subsec:large_debris}), for which the disk remains optically thin and mass is removed primarily by radiation-pressure blowout at micron sizes on sub-Myr timescales. In both cases, collisional evolution followed by dust removal strongly suppresses re-accretion, allowing Mercury to retain the high core mass fraction produced by the initial mantle-stripping impact.

%%%%%%%%%%%%%%%%%%%%%%%%%%%%%%%%%%%%%%%%%%%%%%%%%%
\section{Discussion}\label{sec:discussion}
%%%%%%%%%%%%%%%%%%%%%%%%%%%%%%%%%%%%%%%%%%%%%%%%%%

In this study, we investigated whether collisional evolution can prevent the re-accretion of Mercury's impact ejecta. In the optically thin regime, the debris lifetime is primarily determined by collisional evolution, which governs the supply of dust down to the radiation-pressure blowout limit, where the smallest grains become unbound. In the optically thick regime, most of the debris is shielded from solar radiation, so radiation pressure cannot truncate the cascade. Collisions can therefore continue to break down the debris to nanometre-sized grains, where the collisions transition from destructive fragmentation to predominantly dissipative. The subsequent collisional evolution effectively damps the eccentricity and inclination of the nano-dust, forming a geometrically thin disk. The lifetime of this disk is then controlled by how efficiently material is removed through the exposed surface layers. In both limits, however, the debris is depleted faster than it is re-accreted by Mercury, preventing substantial re-accretion of the stripped mantle. We now discuss the uncertainties in this picture and its broader implications.

% ------------------------------------------------------%
\subsection{Uncertainties in the debris-evolution model}
% ------------------------------------------------------%

In this work, we considered the evolution of debris generated by a mantle-stripping impact on Mercury within the specific dynamical architecture of the present-day Solar System. In reality, the impact would have taken place in an early Solar System, with additional planetesimals, planetary embryos, and debris from other giant impacts \citep[e.g.,][]{Raymond_2006}.
The orbital properties and masses of the proto-terrestrial planets may also have been different from their present-day configuration \citep[e.g.,][]{Walsh_2011}. These factors can influence the detailed evolution of the debris; for example, the presence of a nearby massive embryo could alter the debris dynamics through additional scattering, excitation, and re-accretion, similar to how debris from Earth can be affected by Venus \citep[e.g.,][]{Jackson_Wyatt_2012}. We also adopt a single ejection velocity ($\vinf = 4\,\kms$) chosen to maximise re-accretion, whereas in reality the ejecta are expected to have a distribution of velocities \citep[e.g.,][]{Emsenhuber2024, Watt_2024}. Higher ejection velocities would lead to a more spatially extended debris distribution, reducing the optical depth and potentially modifying the subsequent collisional and dynamical evolution \citep[e.g.,][]{Jackson_2014}.
Another simplification is the assumption of isotropic ejection of debris. The comparison between subsets with different ejection directions in \autoref{sec:res} shows that debris ejected opposite to Mercury's orbital motion are re-accreted more efficiently, whereas debris ejected along the orbital motion spread farther outward and are more likely to be strongly influenced by Venus, with the isotropic case lying between these limiting populations (\autoref{fig:debris_accretion}). However, the resulting variation in re-accreted mass would be small compared with that produced by the range of initial disk masses explored in this study and is therefore unlikely to affect our main conclusions.
Similarly, the two ejecta regimes considered here, large non-vaporized fragments and fully vaporized debris, are idealized cases. A real impact may produce a mixture of large fragments, melt, vapour, and re-condensed droplets, with the subsequent evolution combining aspects of both regimes.

If the ejecta contain both vaporized condensates and large fragments, the two populations are expected to evolve approximately independently because of the large difference in their collisional evolution timescales \citep[see \autoref{eq:tcol}; a similar treatment was also adopted by][]{Jackson_Wyatt_2012}. To first order, the total re-accreted mass can therefore be estimated by combining the re-accreted fractions from the two populations weighted by their mass fractions. For example, if $50\%$ of a $0.1\,\Mearth$ disk is vaporized, the condensate component is cleared on a timescale of $\sim10^3\,$yr (\autoref{eq:temclr}), with only $\lesssim 1\%$ expected to re-accrete, accounting for depletion in addition to the dynamical calculation of \autoref{fig:thin_disk}, while the large-fragment component is expected to re-accrete $\sim 10\%$ by $10\,$Myr (\autoref{fig:coll_depletion}). This gives a total re-accreted fraction of only $\sim5-6\%$ of the initial ejecta mass, which is not expected to significantly affect the final CMF.

We also neglect mutual gravitational interactions between the debris particles. Gravitational scattering and focusing between fragments are expected to be weak when the disk is dynamically excited, as relative velocities of a few $\kms$ (\autoref{fig:local_vrel}) exceed the mutual escape velocities even for the largest fragments ($\sim 0.4\,\kms$ for $D = 600\,\km$). Self-gravity of the disk is also unlikely to drive gravitational instability. A Toomre stability estimate gives $Q \gg 1$ for the dynamically excited large-fragment disk, while for the circularized optically thick disk, $Q$ remains above unity despite significant damping for $e \gtrsim 10^{-5}$ over the disk masses considered here. Self-gravity could nevertheless influence the secular evolution of the disk \citep[e.g.,][]{Sefilian_2021,Sefilian_2023}, although its influence decreases as the disk mass is rapidly depleted.

Our treatment of collisional evolution for large debris in \autoref{subsec:large_debris} uses a statistical approach coupled to $N$-body particles, and does not explicitly follow individual fragmentation or collisional damping. We assume that fragments inherit the orbital properties of their parent bodies. This is suitable while the disk remains dynamically hot \citep{Jankovic_2024}, but may not capture regimes where collisional damping becomes important, such as non-fragmenting collisions between grains near $\Dminlim$ (\autoref{subsec:vaporized_debris}). Furthermore, the treatment of the smallest grains relies on analytical estimates of their charging and removal, which depend on uncertain properties of the early solar wind and radiation environment \citep[e.g.,][]{Wood_2002, Wood_2014, Ribas_2005, Gudel_2007, Johnstone_2015}. These uncertainties affect the exact removal timescales, but not the qualitative result that once small grains are produced, they can be efficiently removed when exposed to solar radiation and wind.

We neglect possible coagulation of grains after the cascade reaches the dissipative regime and relative velocities have been damped \citep[e.g.,][]{Blum_Wurm_2008}. Such growth may occur in the shielded interior of the optically thick disk, where grains may remain neutral or weakly charged, whereas grains exposed to solar UV radiation and solar-wind particles are expected to become charged \citep[e.g.,][]{Mann_2014} and can experience an electrostatic barrier to low-velocity sticking \citep[e.g.,][]{Okuzumi_2009}. Coagulation could therefore modify the detailed size and porosity of the shielded dust, but is unlikely to suppress the surface-layer removal mechanisms unless growth into much larger particles occurs before the grains reach the exposed layers.

Reduced post-impact re-accretion may also affect Mercury's final chemical and isotopic composition. In our canonical large-fragment case, the re-accreted silicate mass is $\sim 45\%$ of the silicate mass remaining in the immediate post-impact Mercury. If proto-Mercury and the impactor had similar initial CMFs, the available silicate mass would be dominated by proto-Mercury because of the larger target mass, although the exact source of the ejecta depends on the impact dynamics. If the two bodies had different chemical or isotopic compositions, the final mantle composition would therefore depend on both the source of the ejecta and the amount that is subsequently re-accreted. For vaporized ejecta, the re-accreted fraction is much smaller, so any compositional signature established during the primary impact is more likely to be preserved. However, even a small amount of re-accreted material could affect the observable surface composition if it forms a detectable veneer to the planet. In particular, if the highly thermally processed ejecta are depleted in volatile elements, limited re-accretion would reduce the contribution of such material to Mercury's surface, consistent with its relatively high abundance of volatile elements \citep[e.g.,][]{McCubbin_2012, Ebel_Stewart_2018}. Quantifying these signatures would require detailed modelling of the chemical and isotopic fractionation during the impact together with the subsequent deposition and mixing of re-accreted material, which is beyond the scope of this work.

Although our results show that re-accretion of the stripped mantle is not an obstacle to the giant-impact hypothesis, other tensions remain. In particular, such a highly energetic event is rare in simulations of terrestrial planet formation \citep[e.g.,][]{Chau_2018, Jackson_2018, Clement_2019, Franco_2022}. Furthermore, MESSENGER observations of volatile abundances on Mercury's surface pose a challenge for scenarios involving highly energetic collisions that could devolatilize the whole planet, although it has been argued that these observations do not rule out a giant-impact origin for Mercury \citep[e.g.,][]{Ebel_Stewart_2018, Spalding_Adams_2020}.

% ------------------------------------------------------%
\subsection{Implications for extrasolar systems}
% ------------------------------------------------------%

Giant impacts have also been proposed to explain some iron-rich rocky exoplanets, often referred to as super-Mercuries \citep[e.g.,][]{Santerne_2018,Bonomo_2019}. However, \citet{Cambioni_2025} find that mantle-stripping giant impacts are unlikely to explain most observed high-density exoplanets. In their simulations, they assume that unresolved debris is re-accreted onto the largest remnant. However, this may underestimate mantle stripping efficiency based on our findings here. The importance of the subsequent debris evolution is also highlighted by \citet{Tajer_2026}, who find that erosive impacts can produce high-CMF planets, but without efficient debris removal, the high CMF is not retained. The question is whether the efficiency of debris removal around super-Mercuries differs substantially from Mercury depending on the impact conditions, planetary masses and stellar environment.

The impact conditions determine 1) the total mass of debris, 2) whether the debris population is dominated by large fragments or small condensates, which sets the optical depth and therefore the dust removal pathway, and 3) the dynamical properties of the debris such as the velocity distribution. The spatial extent of the giant-impact debris disk is primarily set by the ratio of the debris velocity dispersion ($\sigma_v$) to the local Keplerian velocity ($\vk$), with a larger $\sigma_v/\vk$ producing a more extended disk. The debris velocity dispersion generally scales with the progenitor escape velocity \citep{Wyatt_2016,Wyatt_2017}. Debris produced closer to the star is therefore expected to form narrower disks as $\vk$ is larger, while larger $\vesc$ of more massive planets can spread it over a wider region. For a debris velocity distribution scaled to $\vesc$, \citet{Wyatt_2017} estimate the accretion timescale as $t_{\rm acc}\propto M_\star^{-1}a^{4}\rho_{\rm p}^{5/6} M_{\rm p}^{-1/3}$, where $M_\star$ is the stellar mass, and $M_{\rm p}$ and $\rho_{\rm p}$ are the planet mass and density, respectively. This implies significantly faster re-accretion at smaller orbital distances but only a weak dependence on planet mass. The corresponding collisional lifetime has the same dependence on stellar mass and orbital distance, $\tcol\propto M_\star^{-1}a^{4}M_{\rm p}^{-2/9}\rho_{\rm p}^{-1/9}\Dmax\Mdisk^{-1}$ \citep{Wyatt_2017}\footnote{\autoref{eq:tcol} recovers the same $a^4$ scaling when $\Delta a$ and $\Delta h$ are scaled consistently with $\sigma_v/\vk$.}. The fraction of debris that is re-accreted before being collisionally depleted primarily depends on the ratio $t_{\rm acc}/\tcol$, which is independent of orbital distance and stellar mass, and has only a weak dependence on planet mass \citep[e.g.,][]{Wyatt_2017,Ghosh_2025}. Thus, the competition between re-accretion and collisional depletion may remain similar even at much shorter orbital periods, although both processes occur faster. However, the details of the primary giant impact and the influence of neighbouring planets require more detailed consideration.

The efficiency of the different debris-removal mechanisms may also vary between systems. PR drag and stellar-wind drag become faster at smaller orbital distances, with the timescales scaling approximately as $a^2$ (see \autoref{eq:tpr} and \autoref{eq:tsw}), and may therefore be more important for close-in super-Mercuries. Electromagnetic removal may also differ between systems, as both the grain charging and subsequent Lorentz-force evolution depend on the stellar ionizing UV flux, magnetic field, and wind environment.

In general, our results indicate that collisional depletion is likely to play a significant role in removing debris from mantle-stripping collisions even in exoplanetary systems.

A similar collisional evolution of giant-impact debris may help explain the warm dust observed in extrasolar systems. Observations with \emph{Spitzer}, \emph{WISE}, and now \emph{JWST} have revealed variable infrared emission from warm dust, often referred to as extreme debris disks, with changes on year-to-decade timescales consistent with recent giant impacts in close-in systems. 
For example, systems such as ID8 and P1121 show mid-infrared variability consistent with impact-generated dust clouds \citep[e.g.,][]{Meng_2014_ID8, Su_2019_EDD}, while HD~172555 shows spectroscopic evidence for silica-rich dust and the first detection of CO gas associated with giant-impact debris \citep[e.g.,][]{Lisse_2009_HD172555, Schneiderman_2021}, and HD~23514 shows dust and gas signatures consistent with terrestrial-region collisions \citep[e.g.,][]{Su_2025_HD23514}.
Our results suggest that, in the inner regions of planetary systems, the observable signatures of such impact-generated dust may depend sensitively on whether the debris is ejected primarily as large fragments that remain optically thin, or whether a substantial fraction is vaporized by the impact and recondenses into small droplets that form an optically thick debris cloud.
In optically thin disks, the cascade is truncated near the radiation-pressure blowout limit, with the disk lifetime set by the collisional timescale. In optically thick disks, however, collisional evolution can produce a geometrically thin disk of grains smaller than the radiation-pressure blowout limit, and the lifetime is instead determined by the processes controlling dust removal. 
A detailed exploration of optically thick debris evolution in extrasolar systems and its observational implications is left for future work.

%%%%%%%%%%%%%%%%%%%%%%%%%%%%%%%%%%%%%%%%%%%%%%%%%%
\section{Conclusions}\label{sec:conclusions}
%%%%%%%%%%%%%%%%%%%%%%%%%%%%%%%%%%%%%%%%%%%%%%%%%%

In this study, we have explored the fate of impact-generated debris from a mantle-stripping giant impact on Mercury, accounting for both the dynamical and collisional evolution of the ejecta.

We find that collisional evolution efficiently depletes the ejecta, preventing significant re-accretion compared to purely dynamical expectations. In the scenario where the debris is dominated by large, non-vaporized fragments, the disk remains optically thin and evolves through a collisional cascade that produces micron-sized grains on sub-Myr timescales, which are subsequently removed by radiation pressure. In this regime, only a small fraction of the initial mass is ultimately re-accreted, allowing Mercury to maintain the high core mass fraction produced by the giant impact.

If the impact is sufficiently energetic to vaporize the ejecta, the evolution proceeds in a qualitatively different regime. The debris forms an optically thick disk, preventing efficient dust removal by radiation pressure. As a result, the collisional cascade is not truncated at the radiation-pressure blowout limit, but instead extends to much smaller sizes, until impacts are no longer energetic enough to cause further fragmentation. For Mercury's impact ejecta, this transition occurs at nanometre sizes, after which collisions become increasingly dissipative and damp the orbital eccentricities and inclinations, producing an optically thick but geometrically thin disk of nanometre-sized grains. These grains acquire large charge-to-mass ratios through solar UV irradiation in the exposed surface layers and become strongly coupled to the solar magnetic field, enabling rapid removal by electromagnetic forces. The corresponding clearing timescales can be as short as a few thousand years, significantly shorter than the re-accretion timescale.
These results imply that a short-lived population of nanometre-sized dust is a natural outcome of highly energetic giant impacts in the inner Solar System that produce optically thick debris disks.

Overall, our results show that re-accretion of impact ejecta does not pose a fundamental obstacle to the giant impact origin of Mercury’s anomalously large core, and that the coupled dynamical and collisional evolution of debris can play a key role in determining the final densities and compositions of planets growing via giant impacts in the inner Solar System.

%%%%%%%%%%%%%%%%%%%%%%%%%%%%%%%%%%%%%%%%%%%%%%%%%%
%%%%%%%%%%%%%%%%%%%%%%%%%%%%%%%%%%%%%%%%%%%%%%%%%%
\section*{Acknowledgements}

We thank the anonymous referee for insightful comments and suggestions. 
T.G. gratefully acknowledges support from the Leverhulme Centre for Life in the Universe at the University of Cambridge through the Joint Collaborations Research Project Grant G112026, LGAG/393.
A.P.J. gratefully acknowledges support from NASA grant 80NSSC23K0267, and the Fisher Endowed Chair of Biological and Physical Sciences at Towson University.
O.S. acknowledges support from UKRI (STFC) grant UKRI1184 and M.W. from UKRI1198.
We thank Stephen Mojzsis for drawing our attention to this problem and Zoe Leinhardt for a helpful discussion.

%%%%%%%%%%%%%%%%%%%%%%%%%%%%%%%%%%%%%%%%%%%%%%%%%%
\section*{Data Availability}

The data underlying this article will be shared on reasonable request to the corresponding author.

%%%%%%%%%%%%%%%%%%%% REFERENCES %%%%%%%%%%%%%%%%%%

\bibliographystyle{mnras}
\bibliography{ref}

%%%%%%%%%%%%%%%%%%%%%%%%%%%%%%%%%%%%%%%%%%%%%%%%%%

%%%%%%%%%%%%%%%%% APPENDICES %%%%%%%%%%%%%%%%%%%%%
% \appendix
% \section{Appendix A}\label{app:A}
%%%%%%%%%%%%%%%%%%%%%%%%%%%%%%%%%%%%%%%%%%%%%%%%%%

%%%%%%%%%%%%%%%%%%%%%%%%%%%%%%%%%%%%%%%%%%%%%%%%%%
% Don't change these lines
\bsp	% typesetting comment
\label{lastpage}
\end{document}